\documentclass[aps,prd,showpacs,amssymb,nofootinbib,preprintnumbers]{revtex4}
\usepackage{slashed}
\usepackage{subfigure}
\usepackage{amsfonts,graphicx,bm}
\usepackage{amsmath}
\usepackage{rotating}
\usepackage{multirow,tabularx}

\begin{document}

\title{Relatively Large Theta13 from Modification to the Tri-bimaximal, Bimaximal and Democratic Neutrino Mixing Matrices}

\newcommand*{\PKU}{School of Physics and State Key Laboratory of Nuclear Physics and
Technology, \\Peking University, Beijing 100871,
China}\affiliation{\PKU}
\newcommand*{\SDU}{Department of Physics, Shandong University, Jinan, Shandong 250100, China}\affiliation{\SDU}
\newcommand*{\CHEP}{Center for High Energy
Physics, Peking University, Beijing 100871,
China}\affiliation{\CHEP}
\author{Wei Chao}\email{chaow@pku.edu.cn}\affiliation{\CHEP}
\author{Ya-juan Zheng}\email{yjzheng@mail.sdu.edu.cn}\affiliation{\SDU}\affiliation{\PKU}

\begin{abstract}
Inspired by the recent T2K indication of a relatively large
$\theta_{13}$, we provide a systematic study of some general
modifications to three mostly discussed neutrino mixing patterns,
i.e., tri-bimaximal, bimaximal and democratic mixing matrices. The
correlation between $\theta_{13}$ and two large mixing angles are
provided according to each modifications. The phenomenological
predictions of $\theta_{12}$ and $\theta_{23}$ are also discussed.
After the exclusion of several minimal modifications, we still have
reasonable predictions of three mixing angles in $3\sigma$ level for
other scenarios.
\end{abstract}


\maketitle
\section{Introduction}
The observation of neutrino oscillations has revealed that neutrinos
may have non-zero masses and lepton flavors are mixed\cite{pdg}. In
the basis where the flavor eigenstate of three charged leptons  are
identical to their mass eigenstates, the mixing of neutrino flavors
can be described by the standard parameterization\cite{standpara},
which is expressed by three mixing angles $\theta_{12}^{}$,
$\theta_{23}^{}$, $\theta_{13}^{}$ and three CP-violating phases
$\delta, \rho, \sigma$:
\begin{eqnarray}
V = \left( \begin{array}{ccc} c^{}_{12} c^{}_{13} & s^{}_{12}
c^{}_{13} & s^{}_{13} e^{-i\delta} \\ -s^{}_{12} c^{}_{23} -
c^{}_{12} s^{}_{13} s^{}_{23} e^{i\delta} & c^{}_{12} c^{}_{23} -
s^{}_{12} s^{}_{13} s^{}_{23} e^{i\delta} & c^{}_{13} s^{}_{23} \\
s^{}_{12} s^{}_{23} - c^{}_{12} s^{}_{13} c^{}_{23} e^{i\delta} &
-c^{}_{12} s^{}_{23} - s^{}_{12} s^{}_{13} c^{}_{23} e^{i\delta} &
c^{}_{13} c^{}_{23} \end{array} \right) P^{}_\nu \;
,\label{standpara}
\end{eqnarray}
where $c^{}_{ij}\equiv \cos \theta^{}_{ij}$, $s^{}_{ij} \equiv \sin
\theta^{}_{ij}$ and $P_\nu^{} \equiv \{ e^{ip}, e^{i\sigma}, 1\}$ is
a diagonal phase matrix, which is physically relevant if neutrinos
are Majorana particles. It is possible to fit neutrino masses and
mixing parameters from various neutrino oscillation experiments.
From\cite{globalfit}, we have
\begin{eqnarray}
&&\Delta m^2_{21} = 7.59\pm 0.20(^{+0.61}_{-0.69})  \times 10^{-5}
\mbox{eV}^2\;,\nonumber\\
&&\Delta m^2_{31} = -2.36\pm 0.11(\pm0.37) [+2.46\pm0.12(\pm0.37)]
\times 10^{-3} \mbox{eV}^2\;,\nonumber\\
&&\theta_{12} = 34.5\pm 1.0(^{+3.2}_{-2.8})^\circ\;,\;\;\theta_{23}
= 42.8^{+4.7}_{-2.9}(^{+10.7}_{-7.3})^\circ\;,\;\;\theta_{13} =
5.1^{+3.0}_{-3.3}(\le 12.0)^\circ\; , \label{globalfit}
\end{eqnarray}
at $1\sigma(3\sigma)$ level.

However, the latest result from long baseline neutrino oscillation
experiment ${\rm T2K} $ indicates that $\theta_{13}^{}$ is
relatively large.  At the $90\% $ confidence level, the T2K
data\cite{tsk} are
\begin{eqnarray}
&  & 5.0^\circ \lesssim \theta^{}_{13} \lesssim 16.0^\circ ~~~~
({\rm ~NH~}) ,
\nonumber \\
&  & 5.8^\circ \lesssim \theta^{}_{13} \lesssim 17.8^\circ ~~~~
(~{\rm IH}~) , \label{t2kres}
\end{eqnarray}
for a vanishing Dirac CP-violating phase $\delta = 0^\circ$, where
"NH" and "IH" correspond to the normal and inverted neutrino mass
hierarchies respectively. The best fit values are $\theta_{13}^{} =
9.7^\circ$ (NH) and $11.0^\circ$ (IH). There are already some
literatures discussing on this issue \cite{xing,
hezee,zheng,Manst,zhou,Araki,haba,meloni,sterli}.

From the theoretical point of view, there are three types of well
motivated neutrino mixing patterns: tri-bimaximal mixing pattern
(TB) \cite{tri}, bimaximal mixing pattern (BM)\cite{bi}, and
democratic mixing pattern (DC) \cite{minzhu}, which may arise from
some discrete flavor symmetries, such as $A_4^{}$ and $\mu-\tau$
symmetry, or some very special structures of neutrino mass matrices.
The explicit forms of them read as follows:
\begin{eqnarray}
V_{\rm TB} = \left ( \begin{array}{rrr}
\sqrt{2\over 3}&\sqrt{1\over 3}&0\\
-\sqrt{{1\over 6}}&\sqrt{{1\over 3}}&\sqrt{1\over 2}\\
-\sqrt{{1\over 6}}&\sqrt{{1\over 3}}&-\sqrt{{1\over 2}}
\end{array}
\right )\; , \hspace{1cm}
V_{\rm BM}=\left(
\begin{array}{rrr}
\sqrt{1\over 2 } & \sqrt{1\over 2 } & 0 \\
-{1 \over 2 }& {1 \over 2 } & \sqrt{1\over 2 } \\
{1 \over 2 } &  -{ 1 \over 2 } &\sqrt{1\over 2 }
\end{array}\right) \; , \hspace{1cm}
V_{\rm DC} = \left ( \begin{array}{rrr}
\sqrt{\frac{1}{2}}&\sqrt{\frac{1}{2}}&0\\
\sqrt{\frac{1}{6}}&-\sqrt{\frac{1}{6}}&-\sqrt{\frac{2}{3}}\\
-\sqrt{\frac{1}{3}}&\sqrt{\frac{1}{3}}&-\sqrt{\frac{1}{3}}
\end{array}
\right )\;. \label{vtri}
\end{eqnarray}
It is obvious that all these three mixing patterns are not
consistent with the current neutrino mixing data, because they all
have predictions of vanishing $\theta^{}_{13}$. Actually, the paper
\cite{xing} considered possible perturbations to the Democratic
neutrino mixing pattern to get the relatively large
$\theta_{13}^{}$. While the paper \cite{hezee} considered minimal
modifications to the tri-bimaximal mixing pattern to fit the T2K
data.

In this paper, we consider some general modifications to these
neutrino mixing patterns to get appropriate neutrino mixing angles
that may fit the T2K result. Since $U_{\rm PMNS}^{} =U_\ell^\dagger
U_\nu^{}$, where $U_\ell^{}$ and $U_\nu^{}$ are used to diagonalize
the charged lepton mass matrix and neutrino mass matrix
respectively, these modifications may come from the neutrino sector
or from the charged lepton sector, and even both. We will discuss
these cases in detail.

The remaining part of this paper is arranged as follows: In section
II, possible modifications are listed. Section III is devoted to
study their phenomenological results. We summarize in section IV.

\section{ Possible modifications for $V^{}_{\rm TB}$, $V^{}_{\rm BM}$ and $V^{}_{\rm DC}$
}

In this section, we consider possible modifications for $V^{}_{\rm
TB}$, $V^{}_{\rm BM}$ and $V^{}_{\rm DC}$. For a given neutrino
mixing matrix $U$, there are three possible forms of modifications:
$U\cdot X$, $Y \cdot U$ or $ Y \cdot U \cdot X$, where $X$ and $Y$
denote generic perturbation matrix. Notice that the neutrino mixing
matrix, which comes from the mismatch between the diagonalizations
of the neutrino mass matrix and the charged lepton mass matrix, is
given by $U_{\rm PMNS}^{} = V_l^\dagger V_\nu^{}$. $U \cdot X$ may
come from perturbations to the original neutrino mass matrix, which
is obtained from certain flavor symmetry, $Y \cdot U$ may come from
perturbations to the original charged lepton mass matrix, while $Y
\cdot U \cdot X$ may come form perturbations to the both sector.

From the mathematic point of view. $X$ and $Y$ can be expressed as
$X(Y)= V^{}_{23} V^{}_{13} V^{}_{12} $ in general, where
$V^{}_{23}$, $V^{}_{13}$ and $V^{}_{12}$ are given by
\begin{eqnarray}
&&V^{}_{12} = \left (\begin{array}{ccc}
\cos x & \sin x &0\\
-\sin x &\cos x &0\\
0&0&1
\end{array}
\right )\;, \;\; V^{}_{23} = \left (\begin{array}{ccc}
1&0&0\\
0&\cos y  &\sin y ~e^{i \delta}\\
0&-\sin y ~e^{-i \delta}& \cos y
\end{array}\right )\;, \;\;V^{}_{13} = \left ( \begin{array}{ccc}
\cos z &0&\sin z ~ e^{i\delta} \\
0&1&0\\
-\sin z ~e^{-i \delta} &0& \cos z
\end{array}
\right )\;. \label{vb}
\end{eqnarray}
where $x$, $y$, $z$ denote rotation angles and $\delta$ denote
possible CP-violating phases. For simplicity, we may turn off one or
two $V^{}_{ij}$ ($ij =1,2,3$). In this way, we may explore easily
the physical meanings of these perturbation, see
Ref\cite{xing,hezee} for illustration. To be consecutive, we first
classify possible simple perturbations and list them below:
\begin{eqnarray}
&& V^{c1}_{\rm PMNS}= V_{\alpha}^{} \cdot V_{ij}^{} \; , \label{p1}\\
&& V^{c2}_{\rm PMNS}= V_{ij}^{} \cdot  V_{\alpha}^{} \; ,\label{p2} \\
&& V^{c3}_{\rm PMNS}= V_{\alpha}^{} \cdot V_{ij}^{} \cdot V^{}_{kl} \; ,\label{p3} \\
&& V^{c4}_{\rm PMNS} =  V_{ij}^{} \cdot V^{}_{kl} \cdot
V_{\alpha}^{} \; ,\label{p4}
\end{eqnarray}
where $\alpha = {\rm TB, BM~or ~ DC }$ and $(ij), (kl) =(12), (13),
(23)$ respectively. We left $Y \cdot U \cdot X$ case for future
discussion due to its tedious calculation. In the next section, we
will investigate the phenomenology of these perturbations, by
indicating their predication on the most mysterious neutrino mixing
angle $\theta_{13}^{}$. Then, future neutrino oscillation result of
$\theta^{}_{13}$ may also exclude or support these modifications.

\section{Phenomenology}
We investigate in this section phenomenologies of perturbation
scenarios listed in Eqs. \ref{p1}-\ref{p4}. We first investigate the
most simple cases presented Eqs. \ref{p1} and \ref{p2}. The
phenomenologies of scenarios $V^{}_{\rm TB} \cdot V^{}_{ij}$ and
$V^{}_{ij} \cdot V^{}_{\rm TB }$ were already studied in
Ref.\cite{hezee}, such that we will only investigate the other two
cases.

For the $V^{}_{\rm BM} \cdot V^{}_{ij}$ and $V^{}_{ij} \cdot
V^{}_{\rm BM }$ case:
\begin{itemize}
\item $ \mathbf{V^{}_{\rm BM} \cdot V^{}_{12} }:$  We have $\theta^{}_{13}
=0$ and $\theta^{}_{23} < 45^\circ$ in this case.  It's thus
excluded by T2K.

\item $ \mathbf{V^{}_{\rm BM} \cdot V^{}_{13} }:$ We have $\theta^{}_{12} >
45^\circ$ in this case. It's thus excluded.

\item $ \mathbf{V^{}_{\rm BM} \cdot V^{}_{23} }:$
Given $\theta^{}_{12} \subset [31.7^\circ,~ 37.7^\circ]$,
$\theta^{}_{13}$ lies in the range $[26.7^\circ,~ 33.7^\circ]$. It's
thus excluded.

\item $ \mathbf{ V^{}_{23} \cdot V^{}_{\rm BM} :}$ $\theta^{}_{13}
=0$ and $\theta^{}_{12} =45^\circ$. It's thus excluded.

\item  $ \mathbf{ V^{}_{13} \cdot V^{}_{\rm BM} :}$ In this case, we
have
\begin{eqnarray}
&&\tan \theta^{}_{12} = \left| { \sqrt{2} \cos z - \sin z
e^{i\delta}
\over \sqrt{2} \cos z + \sin z e^{i \delta}} \right| \; , \\
&& \sin \theta^{}_{13} = \left| {\sin z \over \sqrt{2}} \right| \;
,\\
&& \tan \theta^{}_{13} = \left| \sec z \right| \; .
\end{eqnarray}
In the left panel of Fig. \ref{fig:jar} we plot $\theta^{}_{12}$,
$\theta^{}_{13}$ and $\theta^{}_{23}$ as functions of $z$.  For
simplicity, we choose $\delta=0$. It's clear that there are enough
parameter space for this scenario to fit both the T2K's result on
$\theta^{}_{13}$ and other neutrino oscillation data. For the case
$\delta \neq 0$, we can derive the Jarlskog of this scenario.
\begin{eqnarray}
J= {\sin 2z (2-\sin^2 z) \over 4 \sqrt{2} (1 + \cos^2 z )}\cdot {
(3+\cos 2 z )^2 - 8 \sin^2 2 z \cos^2 \delta \over (3+ \cos 2 z )^2
+ 8 \sin^2 2 z \cos^2 \delta } \sin \delta
\end{eqnarray}
Given $\theta^{}_{13}=9.7^\circ$, we plot in the right panel of Fig.
\ref{fig:jar}, the Jarlskog as function of $\delta$. We may read
from the figure that $J$ lies in the range $[0,~0.08]$ in this
scenario.

\begin{figure}[h!]
\includegraphics[width=2.81in]{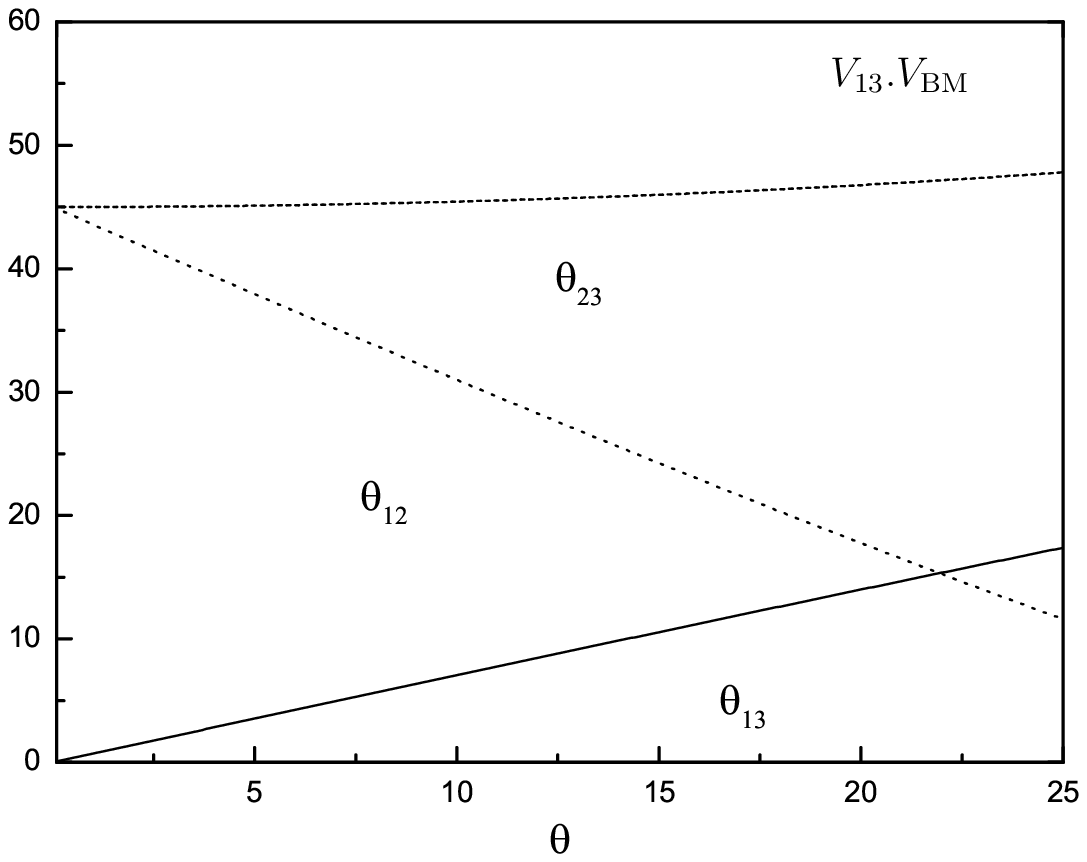}
\includegraphics[width=3.08in]{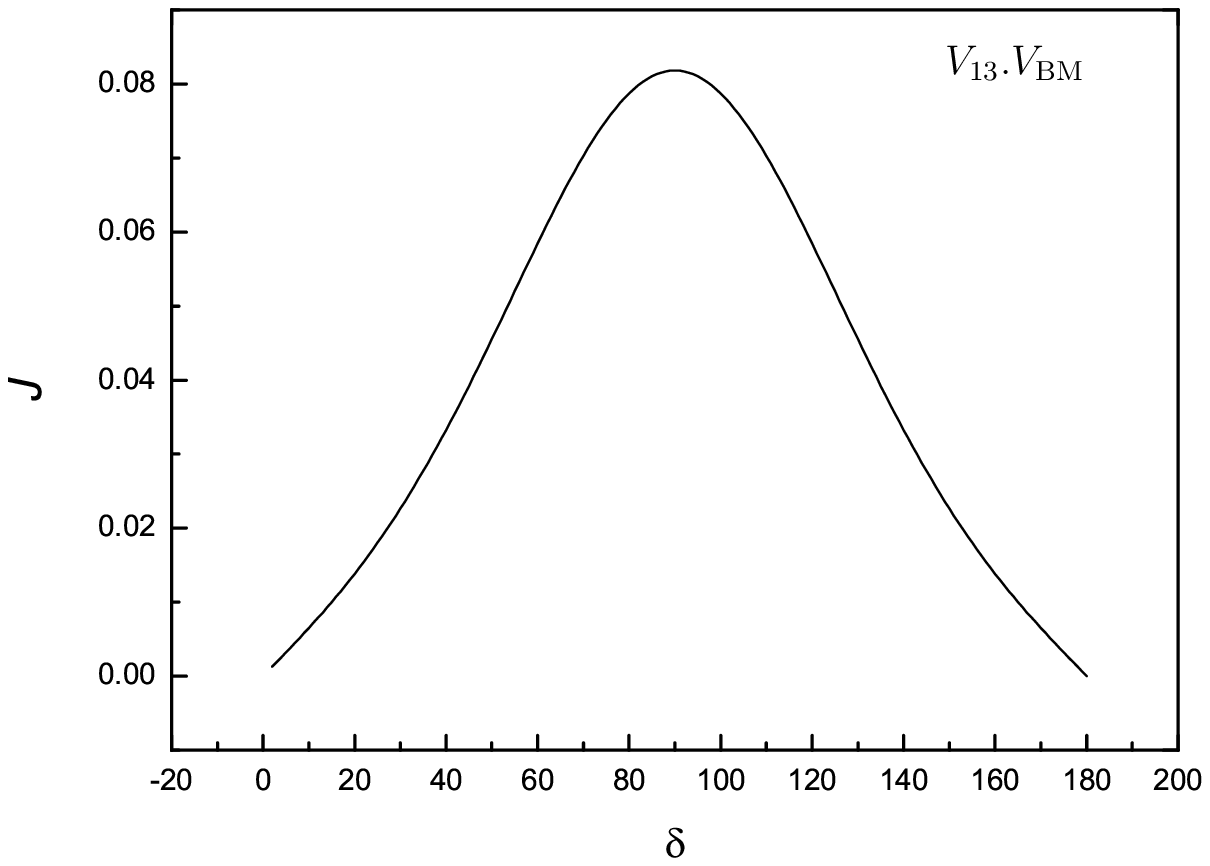}
\caption{Case BM1. $\theta_{ij}$ as a function of $\theta$ (left
panel) and Jarlskog invariant as a function of $\delta$ (right
panel).} \label{fig:jar}
\end{figure}

\item  $ \mathbf{ V^{}_{12} \cdot V^{}_{\rm BM} :}$ We have the
following correlations:
\begin{eqnarray}
&&\tan \theta^{}_{12} = \left| { \sqrt{2} \cos z + \sin z
e^{i\delta}
\over \sqrt{2} \cos z - \sin z e^{i \delta}} \right| \; , \\
&& \sin \theta^{}_{13} = \left| {\sin z \over \sqrt{2}} \right| \;
,\\
&& \tan \theta^{}_{13} = \left| \cos z \right| \; .
\end{eqnarray}
To generate proper $\theta^{}_{12}$, we set $\delta= \pi$. We may
find that this scenario is quite similar to the last one. The only
difference is that this scenario predict $\theta^{}_{23}<45^\circ$,
while the last scenario predict $\theta^{}_{23}>45^\circ$. Both
scenarios are workable.
\end{itemize}

For the $V^{}_{\rm DC} \cdot V^{}_{ij}$ and $V^{}_{ij} \cdot
V^{}_{\rm DC }$ case,
\begin{itemize}
\item $ \mathbf{V^{}_{\rm DC} \cdot V^{}_{12} }:$ $\theta^{}_{13}=0$ in
this case. It's thus excluded.

\item $ \mathbf{V^{}_{\rm DC} \cdot V^{}_{13} }:$ We have $\theta^{}_{12}>
45^\circ$ in this case. It can be excluded.

\item $ \mathbf{V^{}_{\rm DC} \cdot V^{}_{23} }:$ Given $\theta^{}_{12} \subset[31^\circ, ~
38^\circ]$, $\theta^{}_{13}$ lies in the range $[26^\circ,
~35^\circ]$ in this scenario. It's thus excluded.

\item  $ \mathbf{V^{}_{23} \cdot V^{}_{\rm DC}   }:$ $\theta^{}_{13}
=0$. It's excluded.

\item $ \mathbf{V^{}_{13} \cdot V^{}_{\rm DC}   }:$ $\theta^{}_{23} >
54.7^\circ$ in this scenario. It's thus excluded.

\item $ \mathbf{V^{}_{12} \cdot V^{}_{\rm DC}   }:$ In this
scenario, we have $\sin \theta^{}_{13} = \sqrt{(2-\tan^2
\theta^{}_{23})/3}$. Given $\theta^{}_{23}
\subset[35.5^\circ,~53.5^\circ]$,  $\theta^{}_{13}$ lies in the
range $[13.5^\circ, ~45^\circ]$. It's excluded by the best fit value
but allowed by T2K's data. If $\theta^{}_{23} $ is precisely given
in the future neutrino oscillation experiments, We may get to the
denial of this scenario. It would be worth mentioning that proper
$\theta^{}_{12}$ can always be obtained in this case.
\end{itemize}

Now we go to investigate phenomenologies of a little more
complicated cases: $V^{}_{\rm \alpha} \cdot V^{}_{ij}\cdot
V^{}_{kl}$ and $V^{}_{ij} \cdot V^{}_{kl}\cdot V^{}_{\rm \alpha }$
$(\alpha = {\rm BM, TB, DC})$.

\begin{itemize}
\item
$\mathbf{VTBrr1=V_{\rm TB}V_{23}V_{13}:}$ We get the following
correlations
\begin{eqnarray}
\tan \theta_{12}&=&\left|\frac{\cos y}{\sqrt{2}\cos z-\sin y\sin z}\right|,\\
\sin \theta_{13}&=&\left|\frac{1}{\sqrt{3}}(\sqrt{2}\sin z+\cos z\sin y)\right|,\\
\tan \theta_{23}&=&\left|\frac{\cos z(\sqrt{3}\cos y+\sqrt{2}\sin
y)-\sin z}{\cos z(-\sqrt{3}\cos y+\sqrt{2}\sin y)-\sin z}\right|.
\end{eqnarray}
Here $\sin y$ can be expressed as functions of $\theta^{}_{13}$ and
$\theta^{}_{12}$
\begin{eqnarray}
\sin
y=\sqrt{\frac{1-2\tan^2\theta_{12}+3\sin^2\theta_{13}}{1+\tan^2\theta_{12}}}.
\label{TBrr1_siny}
\end{eqnarray}
Then we can get $\sin z$ from the relation of $\theta_{13}$ with $y$
and $z$, i.e.
\begin{eqnarray}
\sin z=\frac{\sqrt{6}\sin\theta_{13}+\sin
y\sqrt{2-3\sin\theta_{13}^2+\sin^2 y}}{2+\sin^2 y}.
\label{TBrr1_sinz}
\end{eqnarray}
Substituting Eq.\ref{TBrr1_siny} and \ref{TBrr1_sinz} into
$\tan\theta_{23}$, we can obtain the relations between $\theta_{23}$
and $\theta_{13}$ which is shown in Fig.\ref{fig:TBrr1}.
\begin{figure}[h!]
\includegraphics[width=3.5in]{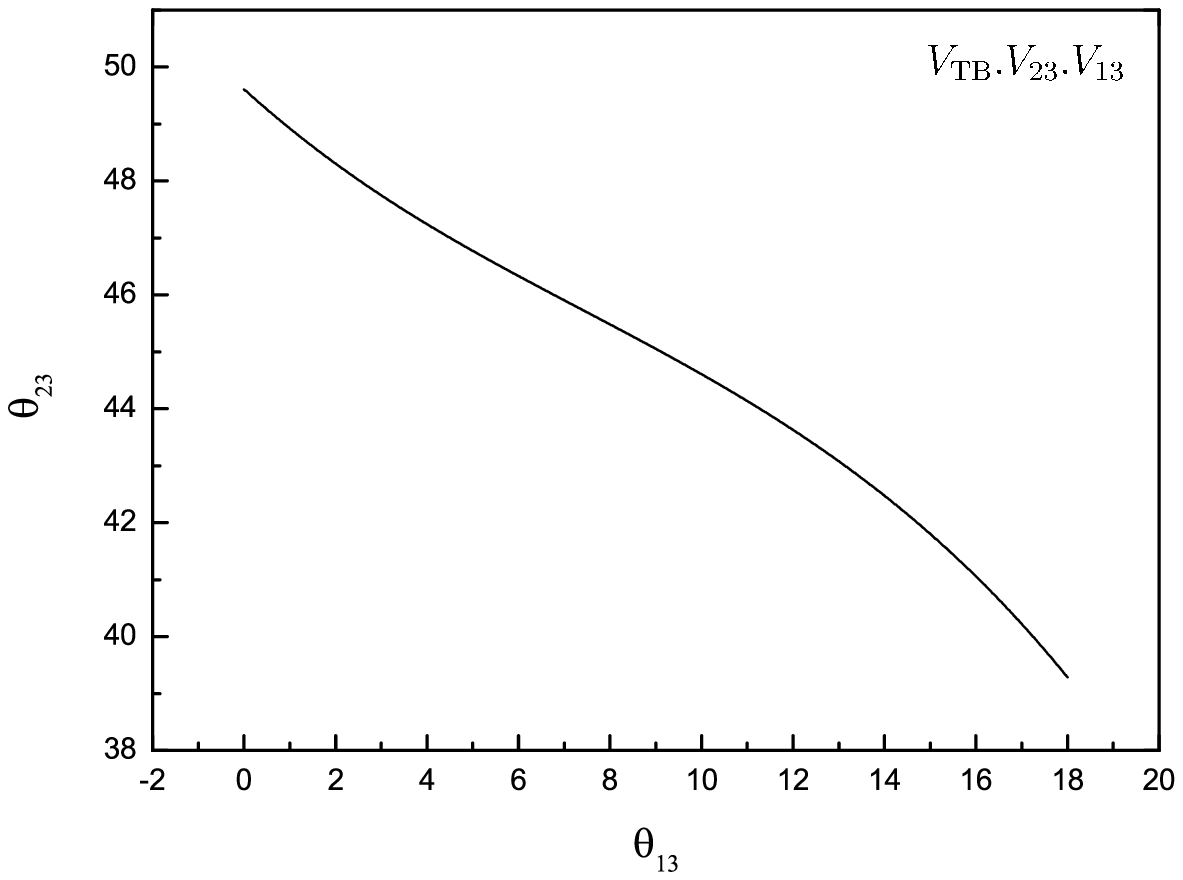}
\caption{Case VTBrr1. $\theta_{23}$ as a function of $\theta_{13}$.}
\label{fig:TBrr1}
\end{figure}
We can see from Fig.~\ref{fig:TBrr1} that as $\theta_{13}$ increases
from $0^\circ$ to $18^\circ$, decreasing $\theta_{23}$ from
$50^\circ$ to $39^\circ$ could be get. Both $\theta_{13}$ and
$\theta_{23}$ are predicted in the experimentally allowed region.
\item
$\mathbf{VTBrr2=V_{\rm TB}V_{23}V_{12}:}$ We have
\begin{eqnarray}
\nonumber
\tan\theta_{12}&=&\left|\frac{\cos x\cos y+\sqrt{2}\sin x}{\sqrt{2}\cos x-\cos y\sin x}\right|,\\
\nonumber
\sin\theta_{13}&=&\left|\frac{1}{\sqrt{3}}\sin y\right|,\\
\tan\theta_{23}&=&\left|\frac{\sqrt{3}\cos y+\sqrt{2}\sin
y}{-\sqrt{3}\cos y+\sqrt{2}\sin y}\right|,
\end{eqnarray}
From the above equations, we could get
\begin{eqnarray}
\sin\theta_{13}=\left|\frac{1-\tan\theta_{23}}{\sqrt{5(1+\tan\theta_{23}^2)-2\tan\theta_{23}}}\right|.
\end{eqnarray}

\begin{figure}[h!]
\includegraphics[width=2.81in]{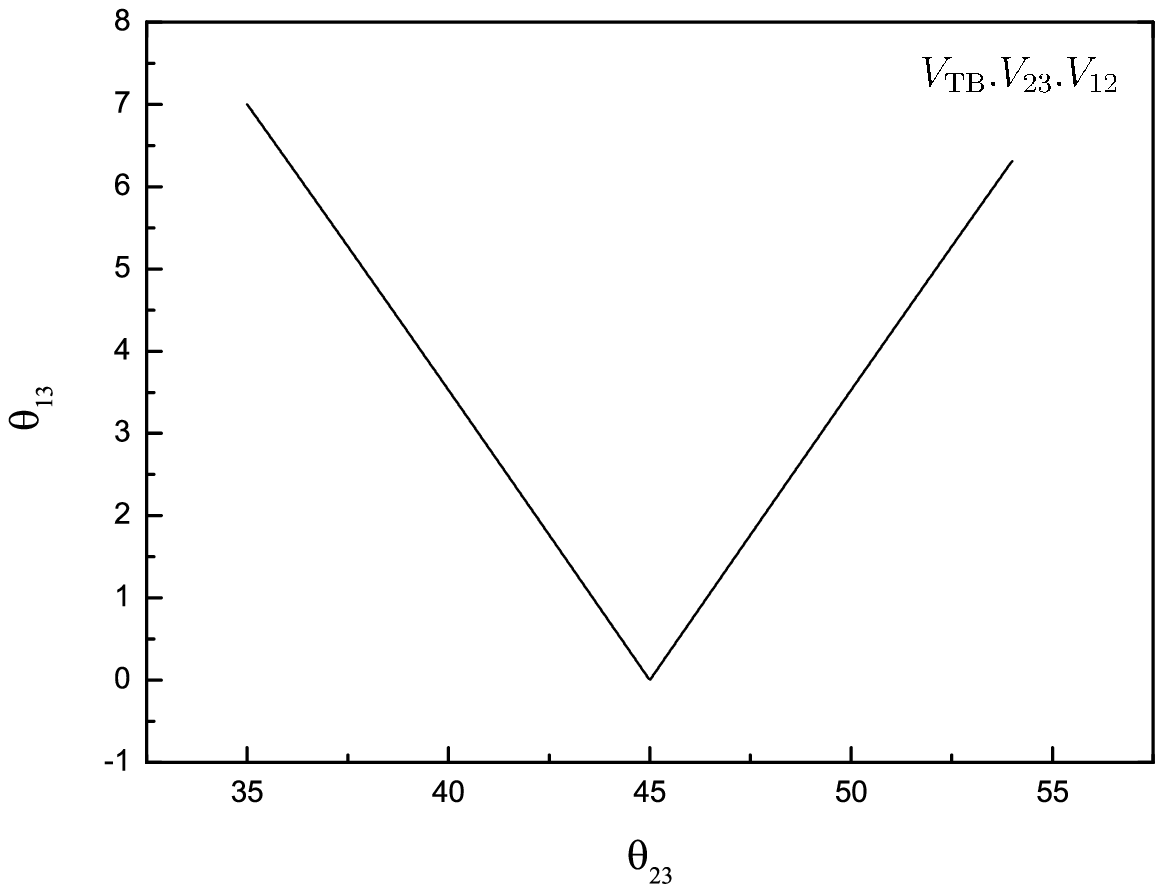}
\includegraphics[width=2.87in]{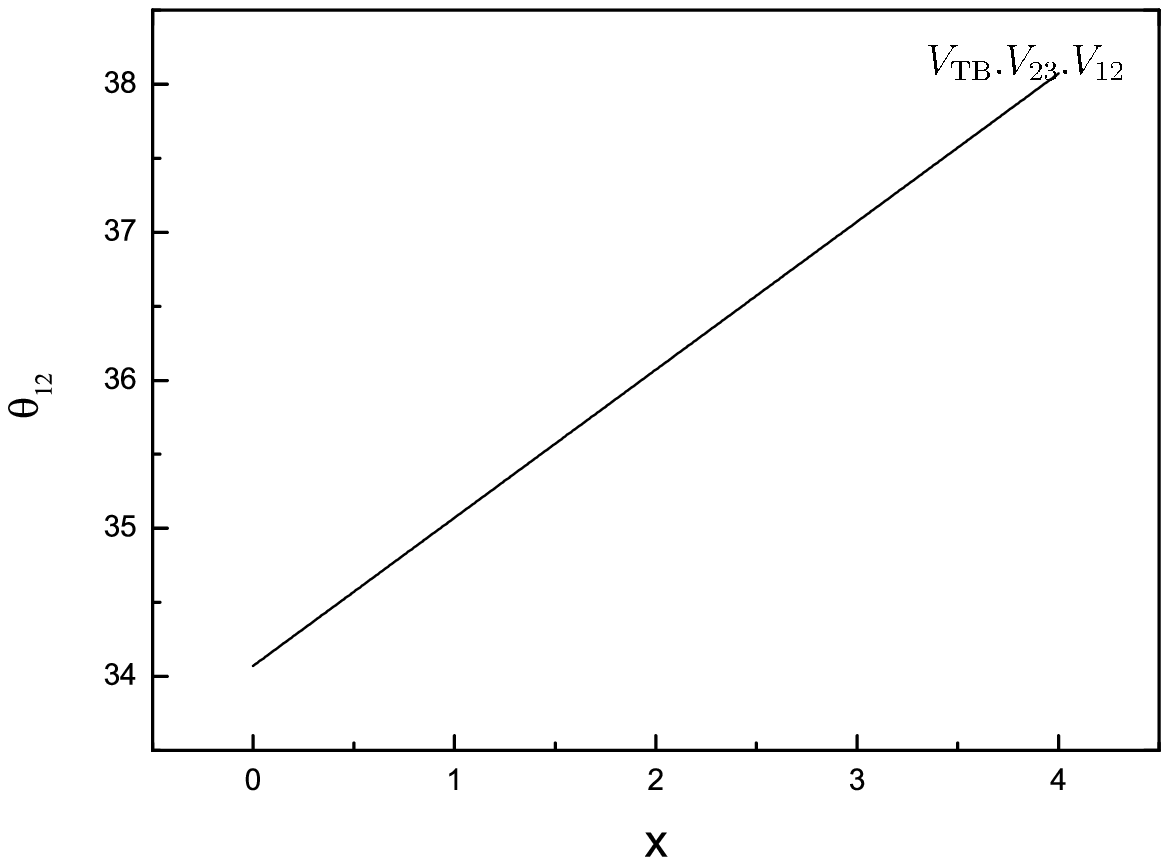}
\caption{Case VTBrr2. $\theta_{13}$ as a function of $\theta_{23}$
(left panel) and $\theta_{12}$ as a function of $x$ (right panel).}
\label{fig:TBrr2}
\end{figure}

The illustration of $\theta_{13}$ as a function of $\theta_{23}$ and $\theta_{12}$ as a function of Euler angle $x$ is shown in Fig.~\ref{fig:TBrr2}. We can see from the left panel that this kind of modification from tribimaximal mixing pattern provides the prediction of $\theta_{13}$ value up to $6^\circ$ while the $\theta_{23}$ varies in the region allowed by current experimental data. In the right panel, reasonable $\theta_{12}$ could also be produced in this case when rotation angle $x$ is changed.   \\
\item
$\mathbf{VTBrr3=V_{\rm TB}V_{13}V_{12}:}$ For this scenario, we have
\begin{eqnarray}
\nonumber
\tan\theta_{12}&=&\left|\frac{\sqrt{2}\cos z\sin x+\cos x}{\sqrt{2}\cos x\cos z-\sin x}\right|,\\
\nonumber
\sin\theta_{13}&=&\left|\sqrt{\frac{2}{3}}\sin z\right|,\\
\tan\theta_{23}&=&\left|\frac{\sqrt{3}\cos z-\sin z}{-\sqrt{3}\cos
z-\sin z}\right|.
\end{eqnarray}
Thus $\theta_{13}$ and $\theta_{23}$ have the following correlation:
\begin{eqnarray}
\sin\theta_{13}=\frac{1-\tan\theta_{23}}{\sqrt{2(1-\tan\theta_{23}+\tan^2\theta_{23})}}.
\end{eqnarray}
The numerical results of this scenario is shown in
Fig.~\ref{fig:TBrr3}
\begin{figure}[h!]
\includegraphics[width=2.81in]{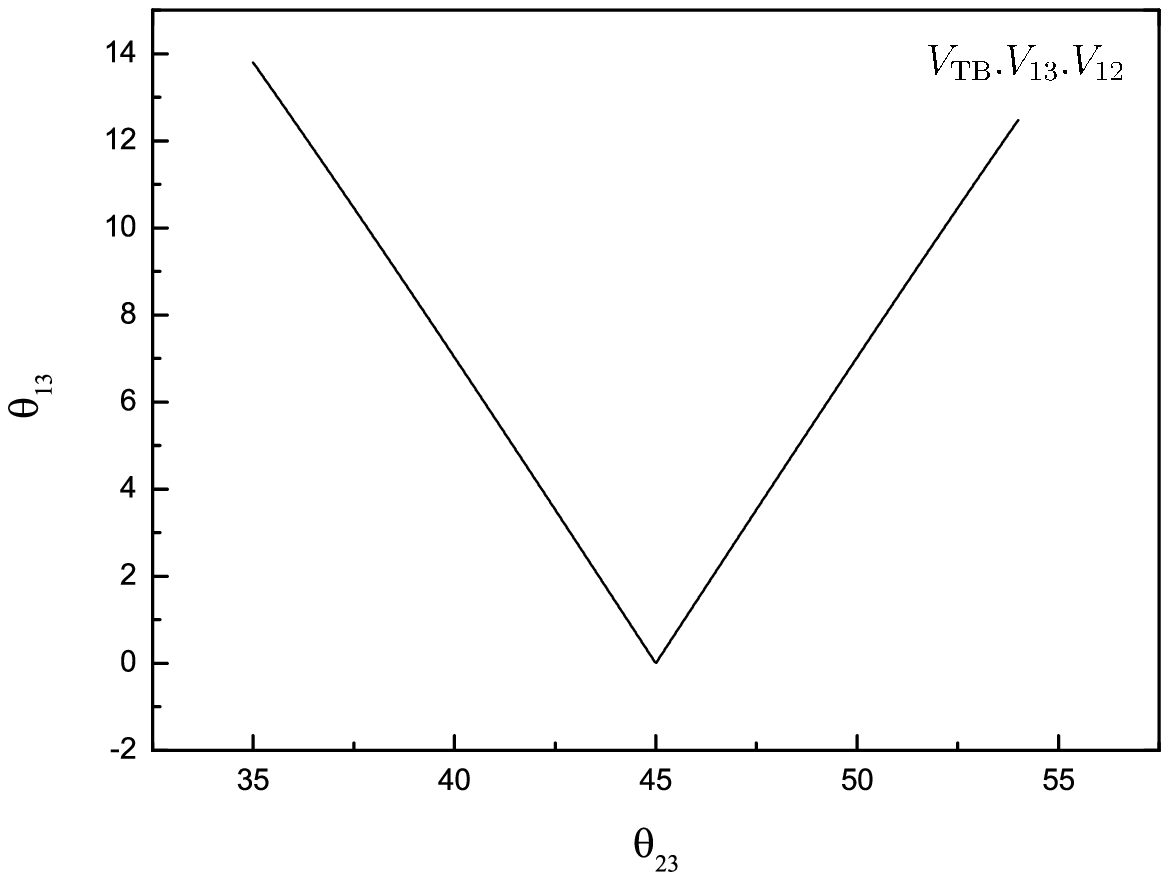}
\includegraphics[width=2.86in]{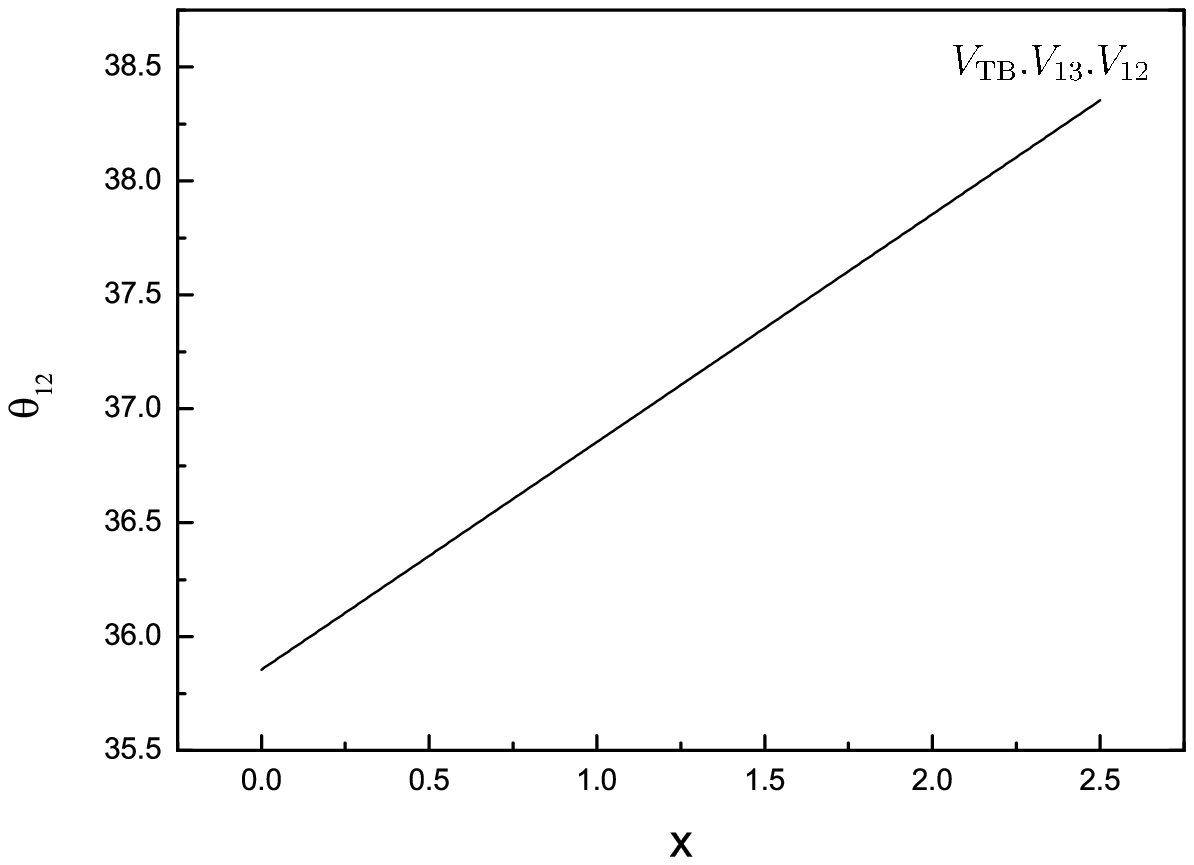}
\caption{Case VTBrr3. $\theta_{13}$ as a function of $\theta_{23}$
(left panel) and $\theta_{12}$ as a function of $x$ (right panel).}
\label{fig:TBrr3}
\end{figure}

In this case,  $\theta_{13}$ varies from $0^\circ$ to almost
$12^\circ$ when $\theta_{23}$ is set in its experimentally allowed
region. In addition, when we  change the corresponding rotaion angle
$x$ from $0^\circ$ to $10^\circ$, $\theta_{12}$ varies from
$35.7^\circ$ to $38.2^\circ$,
 which is consistent with its global fit data in $3\sigma$ level. \\

$\mathbf{VTBll1=V_{23}V_{13}V_{\rm TB}:}$ In this case we have
\begin{eqnarray}
\tan\theta_{12}&=&\left|\frac{\sqrt{2}(\cos z+\sin z)}{2\cos z-\sin z}\right|,\\
\sin\theta_{13}&=&\left|-\sqrt{\frac{1}{2}}\sin z\right|,\\
\tan\theta_{23}&=&\left|\frac{\cos y-\cos z\sin y}{-\cos y\cos
z-\sin y}\right|.
\end{eqnarray}
Then $\theta_{13}$ could be represented by $\theta_{12}$, i.e.
\begin{eqnarray}
\sin\theta_{13}=\frac{|\sqrt{2}\tan\theta_{12}-1|}{\sqrt{4+5\tan^2\theta_{12}-2\sqrt{2}\tan\theta_{12}}}.
\end{eqnarray}

\begin{figure}[h!]
\includegraphics[width=2.81in]{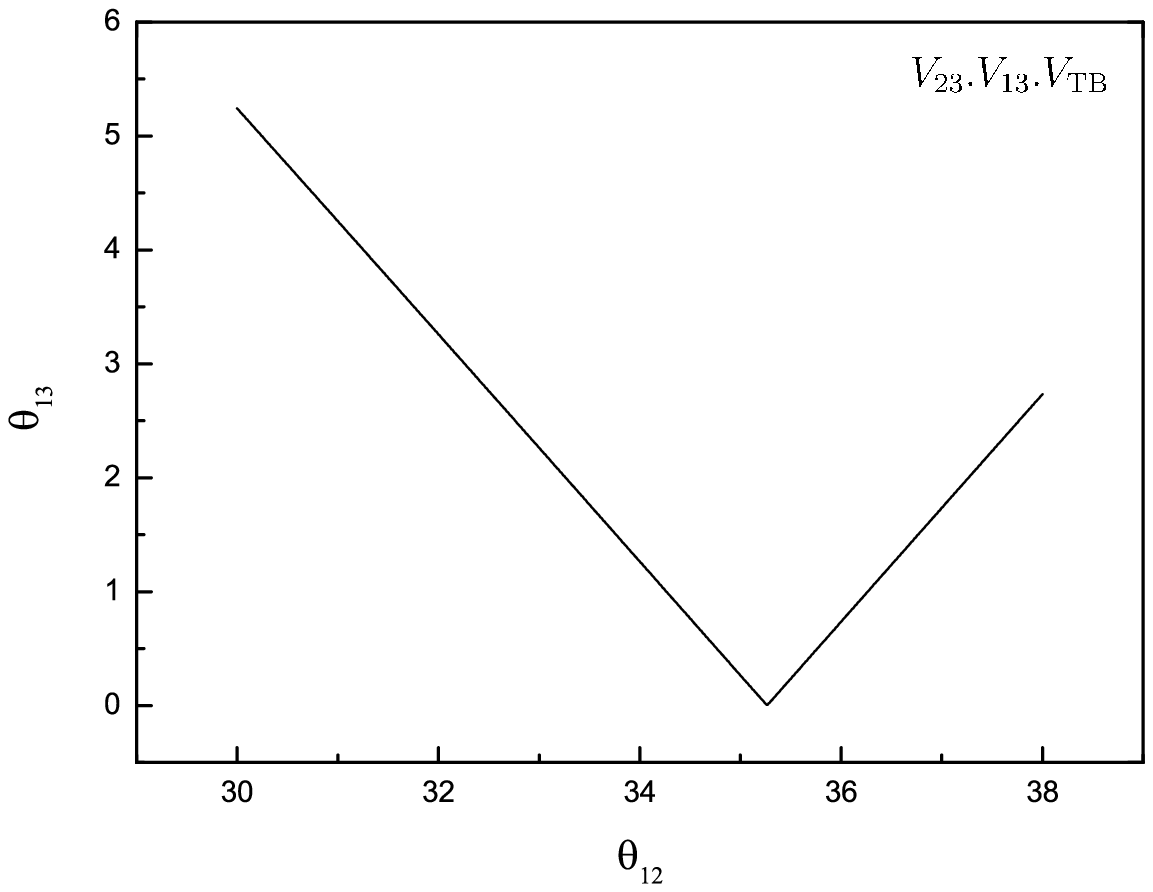}
\includegraphics[width=2.87in]{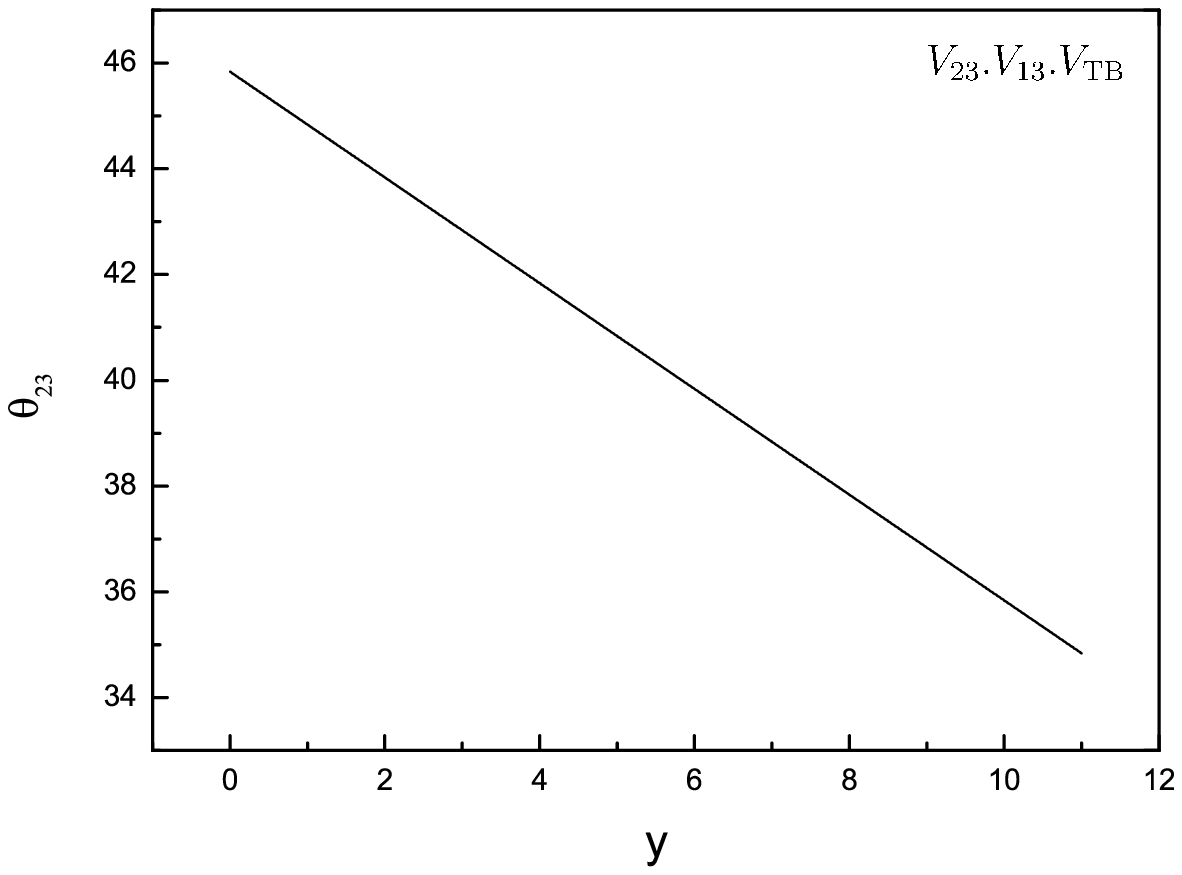}
\caption{Case VTBll1. $\theta_{13}$ as a function of $\theta_{12}$
(left panel) and $\theta_{23}$ as a function of $y$ (right panel).}
\label{fig:TBll1}
\end{figure}
From Fig.~\ref{fig:TBll1}, we can see in the left panel that
$\theta_{13}$ is less than $6^\circ$in this modification scenario from the tribimaximal mixing pattern. In the right panel, $\theta_{23}$ from $34^\circ$ to $46^\circ$ could be produced which lies in the region of global fit data \cite{globalfit} in $3\sigma$ level .\\
\item
$\mathbf{VTBll2=V_{23}V_{12}V_{\rm TB}:}$
\begin{eqnarray}
\nonumber
\tan\theta_{12}&=&\left|\frac{\sqrt{2}(\cos x+\sin x)}{2\cos x-\sin x}\right|,\\
\nonumber
\sin\theta_ {13}&=&\left|\sqrt{\frac{1}{2}}\sin x\right|,\\
\tan\theta_{23}&=&\left|\frac{\cos x\cos y-\sin y}{-\cos y-\cos
x\sin y}\right|,
\end{eqnarray}
Thus, $\theta_{13}$ could be produced, i.e.,
\begin{eqnarray}
\sin\theta_{13}=\frac{\sqrt{2}\tan\theta_{12}-1}{\sqrt{4+5\tan\theta_{12}^2-2\sqrt{2}\tan\theta_{12}}}.
\end{eqnarray}
\begin{figure}[h!]
\includegraphics[width=2.81in]{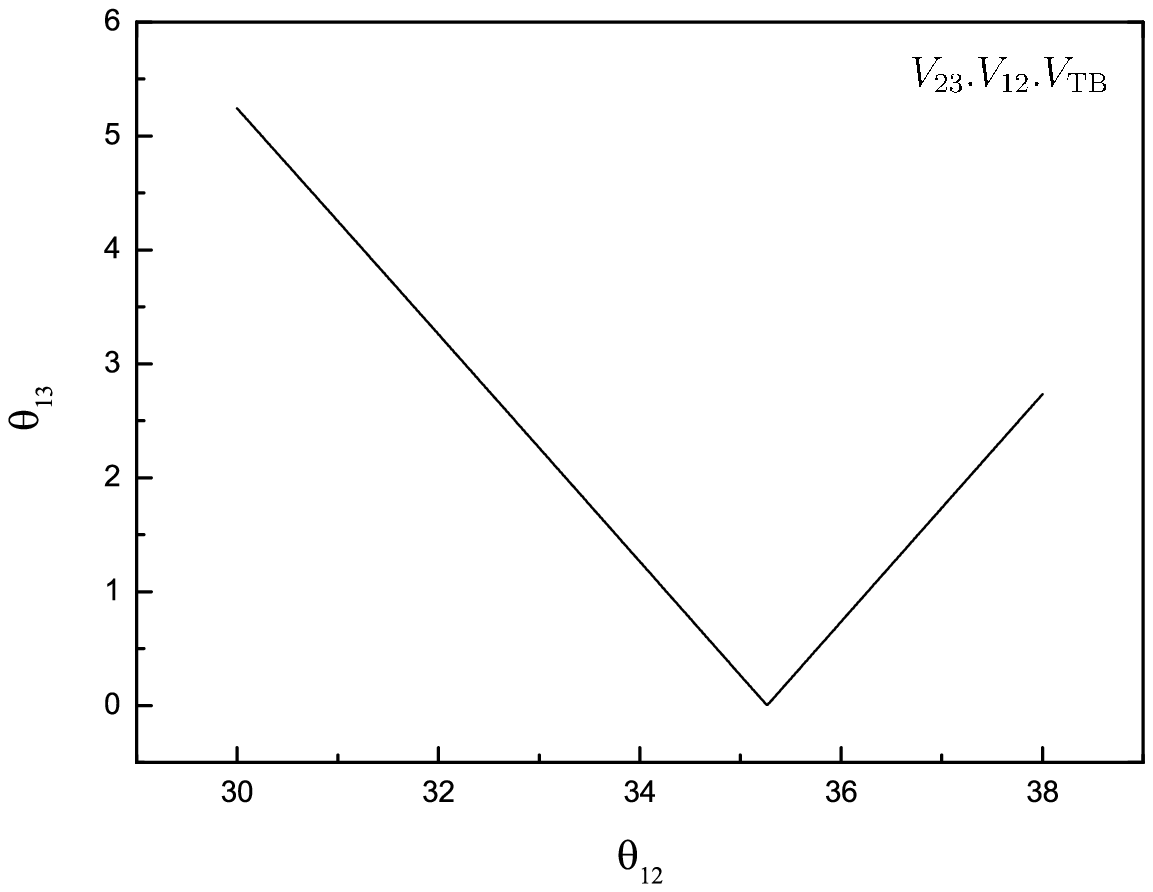}
\includegraphics[width=2.87in]{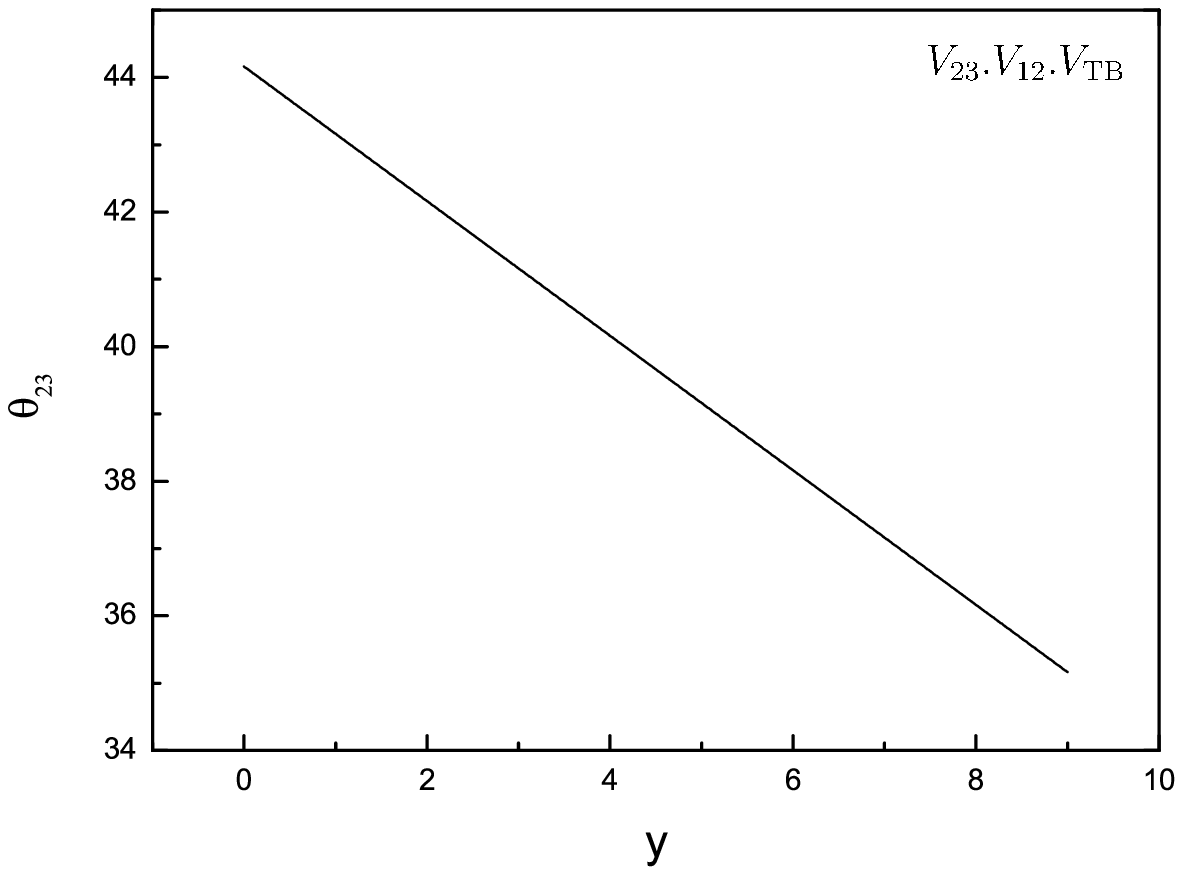}
\caption{Case VTBll2. $\theta_{13}$ as a function of $\theta_{12}$
(left panel) and $\theta_{23}$ as a function of $y$ (right panel).}
\label{fig:TBll2}
\end{figure}
Numerical results are illustrated in Fig.~\ref{fig:TBll2}, it is
obvious that this case shares the same $\theta_{13}$ prediction with
the last case, which could only provides very little $\theta_{13}$,
i.e. less than $6^\circ$. But $\theta_{23}$ evolves differently, the
prediction of this mixing angle could be $44^\circ$ at most, which
is $2^\circ$ smaller than the last scenario.
\item
$\mathbf{VTBll3=V_{13}V_{12}V_{\rm TB}:}$
\begin{eqnarray}
\nonumber
\tan\theta_{12}&=&\left|\frac{\sqrt{2}(\cos x\cos z+\cos z\sin x+\sin z)}{2\cos x\cos z-\cos z\sin x-\sin z}\right|,\\
\nonumber
\sin\theta_{13}&=&\left|\sqrt{\frac{1}{2}}(\cos z\sin x-\sin z)\right|,\\
\tan\theta_{23}&=&\left|\frac{\cos x}{-\cos z-\sin x\sin z}\right|,
\end{eqnarray}
$\cos x$ could be deduced from the above equations, i.e.,
\begin{eqnarray}
\cos
x=\sqrt{\frac{2\tan^2\theta_{23}(1-2\sin^2\theta_{13})}{1+\tan^2\theta_{23}}},
\end{eqnarray}
so is the $\cos z$:
\begin{eqnarray}
\cos z=\frac{\sqrt{2}\sin x\sin\theta_{13}-\sqrt{\sin^2
x-2\sin^2\theta_{13}-1}}{\sin^2 x+1}
\end{eqnarray}
\begin{figure}[h!]
\includegraphics[width=3.3in]{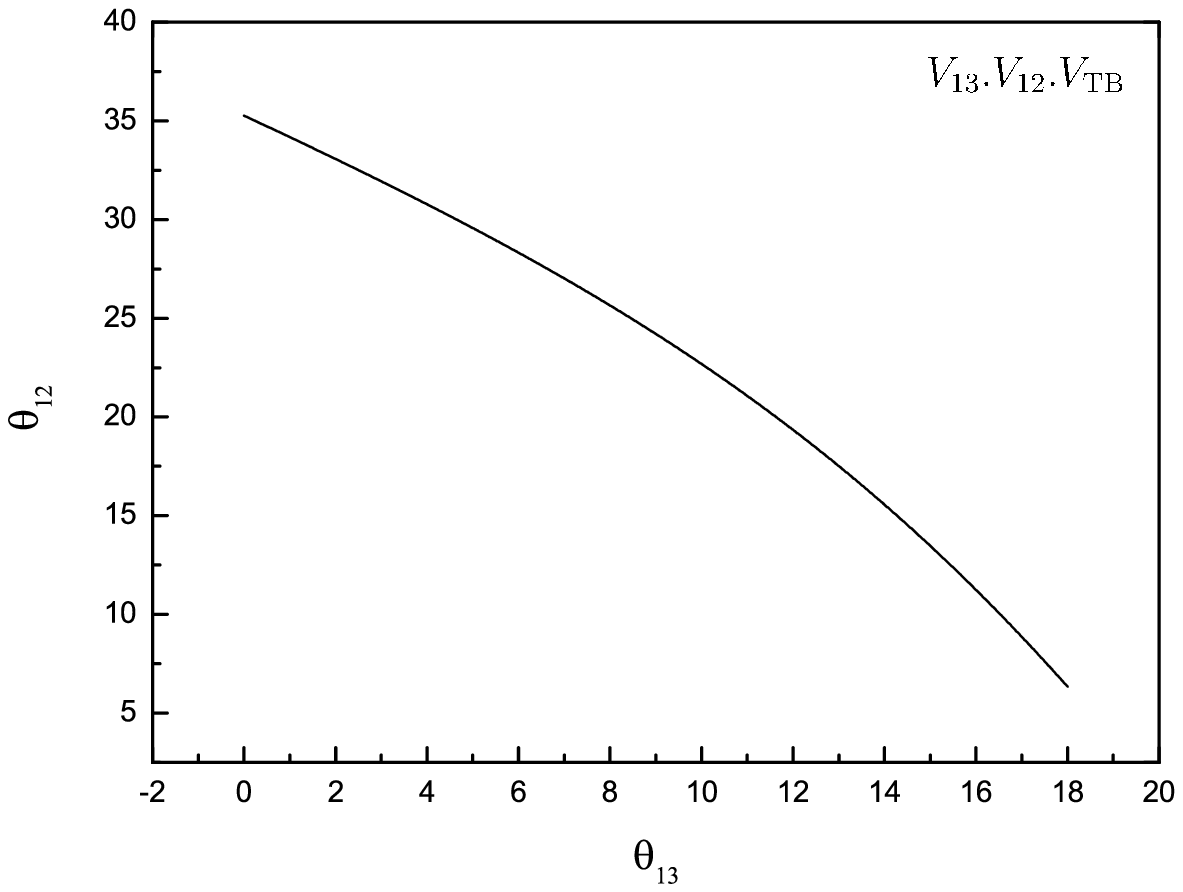}
\caption{Case VTBll3. $\theta_{12}$ as a function of $\theta_{13}$.}
\label{fig:TBll3}
\end{figure}
Numerical results is shown in Fig~\ref{fig:TBll3}. In this case, as
$\theta_{13}$ increases from $0^\circ$ to $18^\circ$ we can get
$\theta_{12}$ from $35^\circ$ to $5^\circ$. It is obvious that
$\theta_{13}$ less than $5^\circ$ is favored if we choose only
reasonable $\theta_{12}$ from $30^\circ$ to $35^\circ$ in the
$3\sigma$ region.

\end{itemize}

For the scenarios $V^{}_{\rm BM} \cdot V^{}_{ij}\cdot V^{}_{kl}$ and
$V^{}_{ij} \cdot V^{}_{kl}\cdot V^{}_{\rm BM }$, we have

\begin{itemize}
\item

$\mathbf{VBMrr1=V_{BM}^{} \cdot V_{13}^{} \cdot V_{12}^{} }:$  For
this case, we obtain
\begin{eqnarray}
&&\tan \theta^{}_{12} = \left|{ \cos x - \cos z \sin x \over \cos x
\cos z
- \sin x} \right| \; , \\
&& \sin \theta^{}_{13} = \left|{ \sin z \over \sqrt{2}}\right| \; , \\
&& \tan \theta^{}_{23} = \left|{\sqrt{2}\cos z - \sin z \over
\sqrt{2} \cos z + \sin z} \right|\; .
\end{eqnarray}
Since both $\theta^{}_{13}$ and $\theta^{}_{23} $depend on a single
parameter $y$, they have the following correlation:
\begin{eqnarray}
\sin \theta^{}_{13} = {1 - \tan \theta^{}_{23} \over \sqrt{3 - 2
\tan \theta^{}_{23} + 3 \tan^2 \theta^{}_{23}}} \; . \label{vb11}
\end{eqnarray}
\begin{figure}[h!]
\includegraphics[width=2.81in]{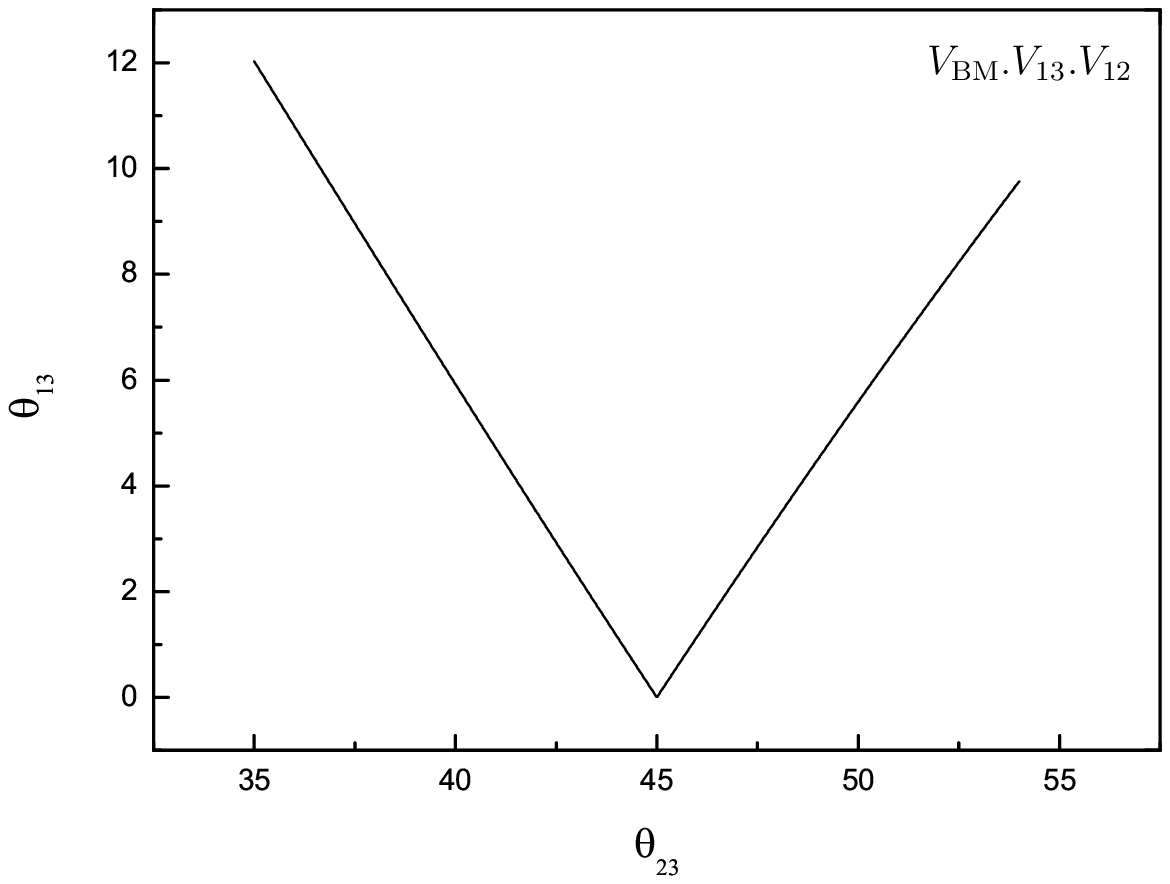}
\includegraphics[width=2.87in]{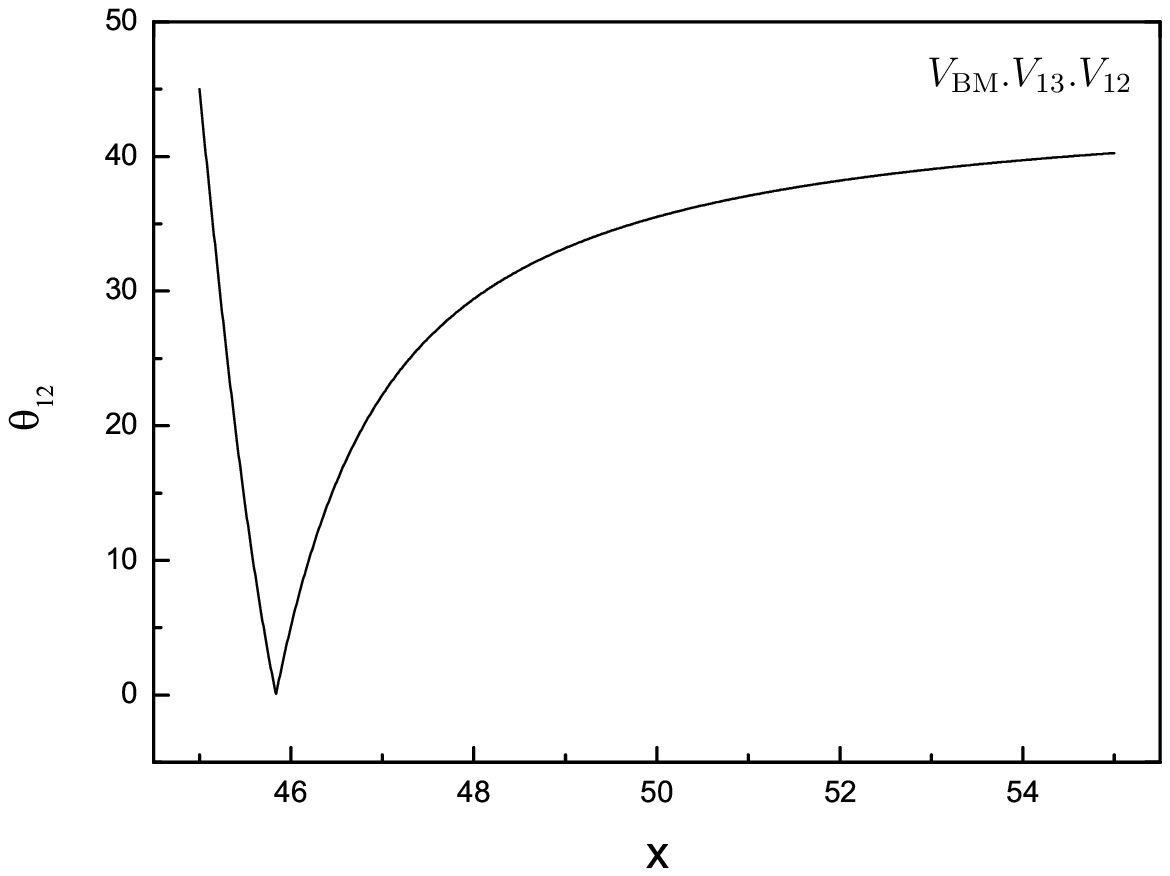}
\caption{Case BMrr1. $\theta_{13}$ as a function of $\theta_{23}$
(left panel) and $\theta_{12}$ as a function of $y$ (right panel).}
\label{fig:BMrr1}
\end{figure}
In the left panel of Fig. \ref{fig:BMrr1}, we plot $\theta^{}_{13}$
as a function of $\theta^{}_{23}$. Taking into account the $3\sigma$
constraint on $\theta^{}_{23}$, the $\theta^{}_{13}$ predicted by
this kind of modification lies in the range $[0^\circ,~ 9.5^\circ]$,
the upper bound of which is approach to the T2K's best fit value for
normal hierarchy case. Appropriate value of $\theta^{}_{12}$ may be
obtained by varying $y$, which is shown in the right panel of Fig.
\ref{fig:BMrr1}.

\item $\mathbf{VBMrr2= V_{\rm BM}^{} \cdot V_{23}^{} \cdot V_{12}^{}:}$  In
this case, we have
\begin{eqnarray}
&&\tan \theta^{}_{12} = \left|{\cos x \cos y - \sin x \over \cos x -
\cos y
\sin x }\right| \; , \\
&&\sin \theta^{}_{13} = \left|{ \sin y \over \sqrt{2}}\right| \; , \\
&& \tan \theta^{}_{23} = \left|{ \sqrt{2} \cos y + \sin y \over
\sqrt{2} \cos y - \sin y }\right| \; .
\end{eqnarray}
It's easy to check that the correlation between $\theta^{}_{13}$ and
$\theta^{}_{23}$ is the same as that of Eq. \ref{vb11}. The only
difference between this scenario and the last one is that their
$\theta^{}_{12}$ evolve differently. Both scenarios predict
reasonable $\theta^{}_{12}$  with a big rotation angle $x$.

\item  $\mathbf{VBMrr3= V_{\rm BM}^{} \cdot V_{23}^{} \cdot
V_{13}^{}}: $ For this case, we derive the following a little
complicated correlations
\begin{eqnarray}
&& \tan \theta^{}_{12} = \left|{ \cos y \over \cos z  - \sin y \sin
z } \right|\; ,
\\
&& \sin \theta^{}_{13} = \left|{ \cos z \sin y + \sin z \over
\sqrt{2} }
\right|\; , \\
&& \tan \theta^{}_{23} =\left| { \sqrt{2} \cos y \cos z + \cos z
\sin y - \sin z \over \sqrt{2} \cos y \cos z - \cos z \sin y + \sin
z } \right|\; .
\end{eqnarray}
We may express $\sin y$ and $\sin z $ as function of
$\theta^{}_{ij}$:
\begin{eqnarray}
&& \sin y = \sqrt{ \cot^2 \theta^{}_{12} + 2 \sin^2
\theta^{}_{13} -1 \over 1 + \cot^2 \theta^{}_{12}} \; , \\
&& \sin z = { \sin \theta^{}_{13} (1 + \tan \theta^{}_{23}) + \tan
\theta^{}_{12} (1 - \tan \theta^{}_{23}) \over \sqrt{2} [1 + \tan
\theta^{}_{23}+ \sin \theta^{}_{13} \tan \theta^{}_{12} (1 - \tan
\theta^{}_{23})]} \; .
\end{eqnarray}
\begin{figure}[h!]
\includegraphics[width=3.5in]{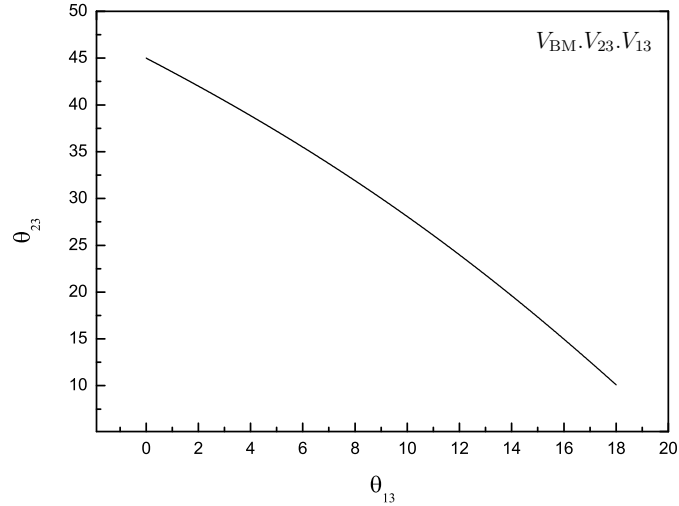}
\caption{Case VBMrr3. $\theta_{23}$ as a function of $\theta_{13}$
.} \label{fig:BMrr3}
\end{figure}
The relationship between $\theta^{}_{ij}$ is a little more
complicated. Here, we only show their numerical results. In the left
panel of Fig. \ref{fig:BMrr3}, we plot $\theta^{}_{23}$ as function
of $\theta^{}_{13}$, choosing the best fit value of
$\theta^{}_{12}$. We may read from the figure that the upper bound
for the $\theta^{}_{13}$ is $6.1^\circ$ for this case.

\item

$\mathbf{VBM\ell\ell1=  V_{13}^{} \cdot V_{12}^{} \cdot V_{\rm
BM}^{}}:$ ~~ It has the following correlations:
\begin{eqnarray}
&& \tan \theta^{}_{12} = \left| {- \sqrt{2} \cos x \cos z + \cos z
\sin x + \sin z \over \sqrt{2} \cos x \cos z + \cos z \sin x + \sin
z }\right| \; ,
\\
&& \sin \theta^{}_{13} =\left|{ - \cos z \sin x + \sin z \over
\sqrt{2}}
\right|\; , \\
&& \tan \theta^{}_{23} = \left|{ \cos x \over \cos z + \sin z \sin x
} \right|\; .
\end{eqnarray}
Here $x$ and $z$ can be expressed as function of $\theta^{}_{ij}$
\begin{eqnarray}
&&\sin x = \sqrt{ {2 \sin^2 \theta^{}_{13} -1 + 2 \cot^2
\theta^{}_{23} \over 1 + \cot^2 \theta^{}_{23}}} \; , \\
&& \sin z = { \sin \theta^{}_{13} (1 - \tan \theta^{}_{12}) + \tan
\theta^{}_{23} (1 + \tan \theta^{}_{12}) \over \sqrt{2} [1 - \tan
\theta^{}_{12} + \sin \theta^{}_{13} \tan \theta^{}_{23} ( 1 + \tan
\theta^{}_{12})]} \; .
\end{eqnarray}
\begin{figure}[h!]
\includegraphics[width=3.5in]{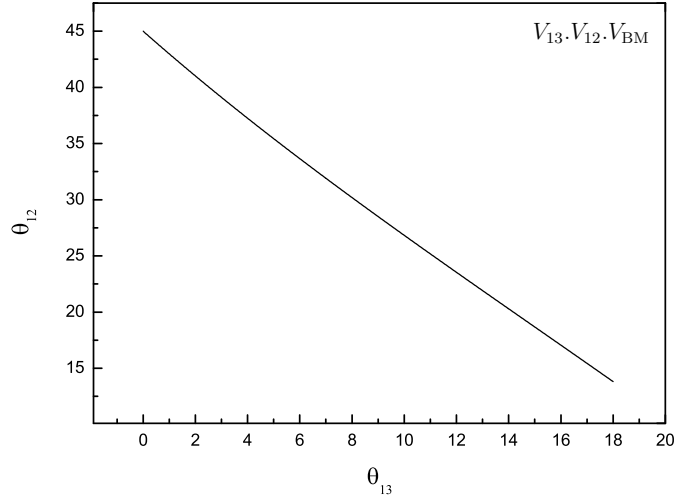}
\caption{Case BMll1. $\theta_{12}$ as a function of $\theta_{13}$.}
\label{fig:BMll1}
\end{figure}
We show in Fig. \ref{fig:BMll1},  $\theta^{}_{12}$ as function of
$\theta^{}_{13}$, setting $\theta^{}_{23}$ its best fit value. We
may find that $\theta^{}_{13}$ lies in the range $[3.7^\circ, ~
7.2^\circ]$, given $\theta^{}_{12}\subset[31.5^\circ,~37.5^\circ]$.
It's impossible to generate larger $\theta^{}_{13}$ in this
scenario.

\item $\mathbf{VBM\ell\ell2=  V_{23}^{} \cdot V_{13}^{} \cdot V_{\rm
BM}^{}}:$ We may derive the following relationships from this
scenario
\begin{eqnarray}
&&\tan \theta^{}_{12}=\left|\frac{\cos z -\sin z }{\cos z + \sin z }\right|\; , \\
&& \sin \theta^{}_{13}= \left|\sqrt{\frac{1}{2}}\sin z \right|,\\
&& \tan \theta^{}_{23}=\left|\frac{\cos y + \cos z  \sin y }{\cos y
\cos z -\sin y }\right|,
\end{eqnarray}
It's easy to check that $\theta^{}_{13}$ and $\theta^{}_{12} $ have
the following correlation:
\begin{eqnarray}
\sin \theta^{}_{13}=\frac{1-\tan \theta^{}_{12}}{2\sqrt{1+\tan
\theta_{12}^2}}
\end{eqnarray}
Since the value $\theta^{}_{13}$ only depend on $\theta^{}_{12}$, we
may constraint the $\theta^{}_{13}$'s parameter space, according to
this equation.  It's numerical result is shown in Fig.
\ref{fig:BMll2}. We may read from this figure that $\theta^{}_{13}$
lies in the range $[6.4^\circ, ~ 10.6^\circ]$.  We plot in the right
panel of Fig. \ref{fig:BMll2} $\theta^{}_{23}$ as function of $y$.
Appropriate $\theta^{}_{23}$ can be obtained by small perturbation
to $y$.

\begin{figure}[h!]
\includegraphics[width=2.81in]{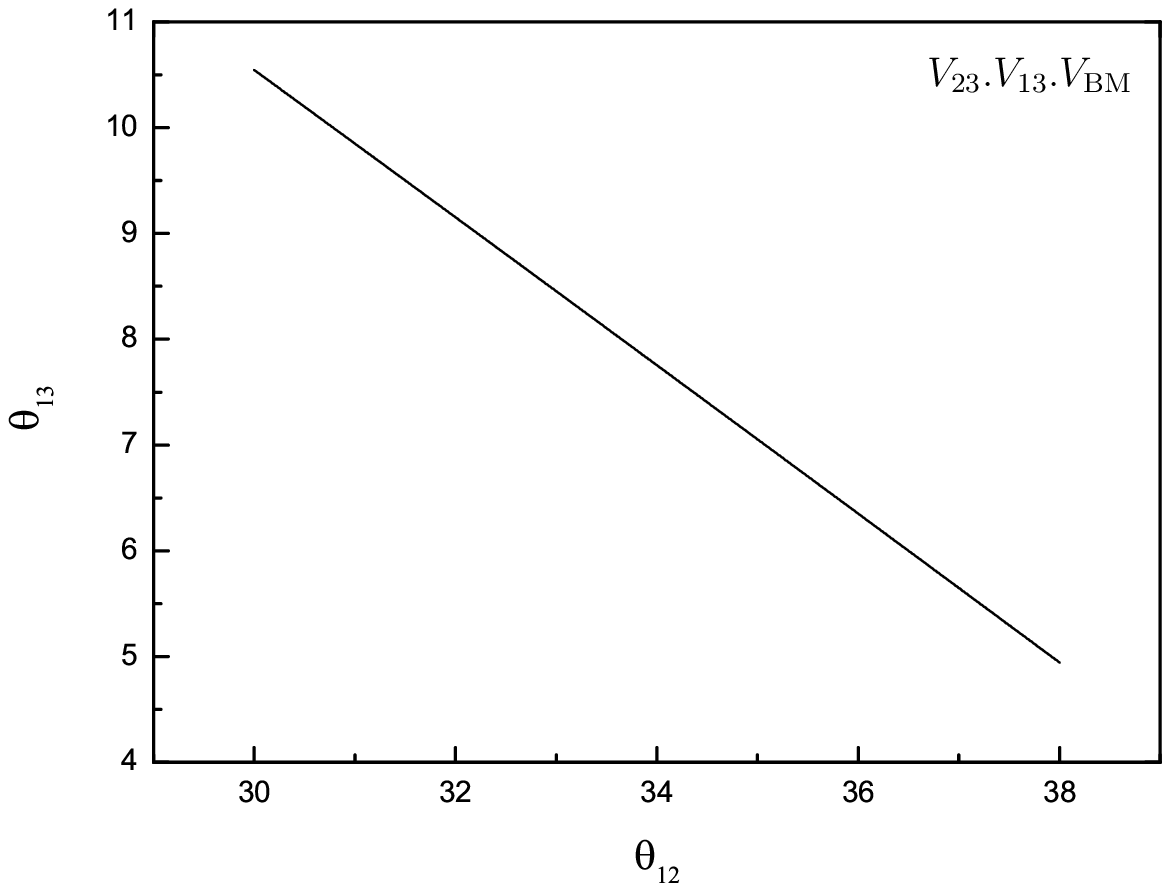}
\includegraphics[width=2.81in]{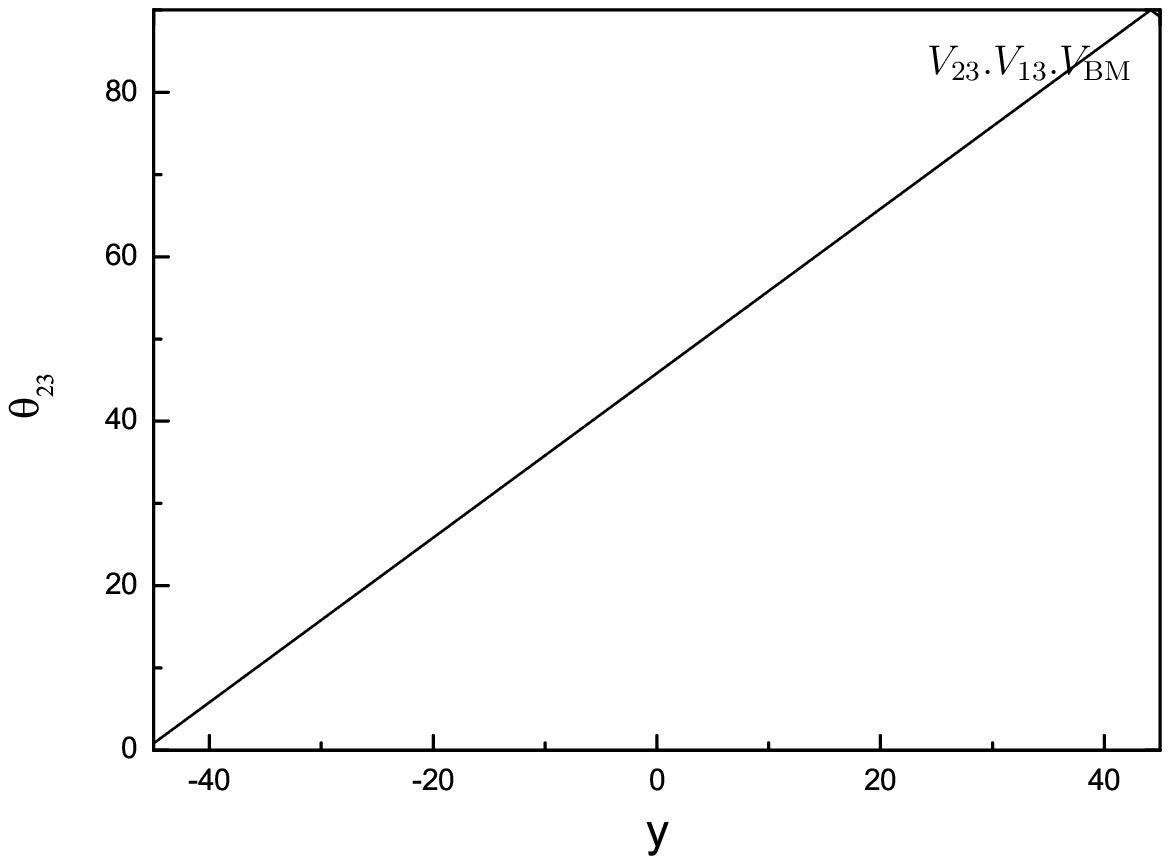}
\caption{Case BMll2. $\theta_{13}$ as a function of $\theta_{12}$
(left panel) and $\theta_{23}$ as a function of $y$ (right panel).}
\label{fig:BMll2}
\end{figure}

\item
$\mathbf{VBM\ell\ell3=  V_{23}^{} \cdot V_{12}^{} \cdot V_{\rm
BM}^{}}:$ The following equations arise in this scenario
\begin{eqnarray}
\nonumber
\tan \theta^{}_{12}&=&\left|\frac{\cos x +\sin x }{\cos x - \sin x }\right|\; ,\\
\sin \theta^{}_{13}&=&\left|\sqrt{\frac{1}{2}}\sin x \right|\;  ,\\
\tan \theta^{}_{23}&=&\left|\frac{\cos x \cos y + \sin y }{\cos y
-\cos x \sin y }\right| \; ,
\end{eqnarray}
where $\theta^{}_{12}$ and $\theta^{}_{13}$ has the following
correlation:
\begin{eqnarray}
\sin \theta^{}_{13}=\frac{1-\tan \theta^{}_{12}}{2\sqrt{\tan
\theta^{}_{12}}}
\end{eqnarray}
\begin{figure}[h!]
\includegraphics[width=2.81in]{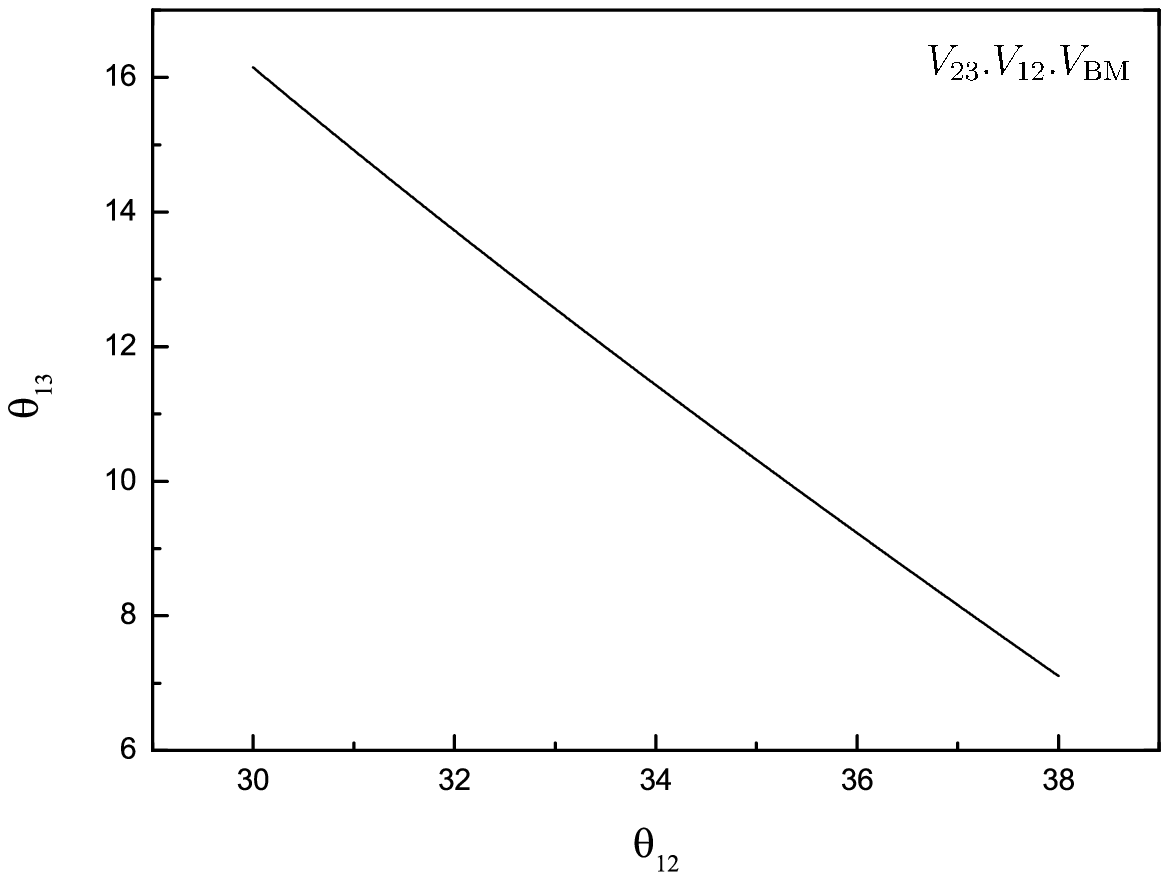}
\includegraphics[width=2.81in]{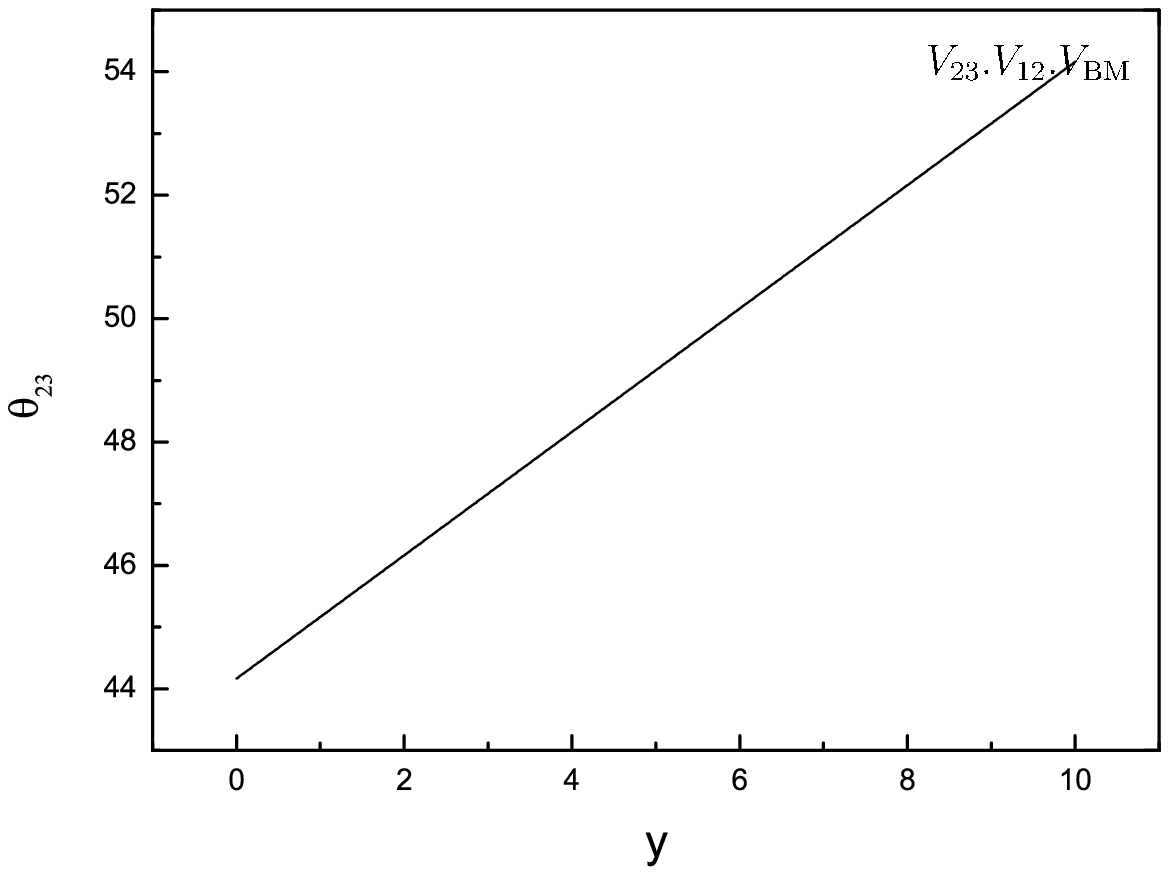}
\caption{Case VBMll3. $\theta_{13}$ as a function of $\theta_{12}$
(left panel) and $\theta_{23}$ as a function of Euler angle y (right
panel).} \label{fig:BMll3}
\end{figure}
As shown in Fig.\ref{fig:BMll3}, $\theta_{13}$ decreases with the
increase of $\theta_{12}$ in this case. When $\theta_{12}$ is set to
lie in its $3\sigma$ global fit region, $\theta_{13}$ ranging from
$8^\circ$ to $15^\circ$ could be predicted which lies in the larger
part of the T2K allowed region. When Euler angle $y$ varies,
$\theta_{23}$ changes from $44^\circ$ to $54^\circ$ which is
consistent with its $3\sigma$ global fit data.

\end{itemize}

For the scenarios $V^{}_{\rm BM} \cdot V^{}_{ij}\cdot V^{}_{kl}$ and
$V^{}_{ij} \cdot V^{}_{kl}\cdot V^{}_{\rm BM }$, we have
\begin{itemize}
\item

$\mathbf{VDCrr1=  V_{\rm DC}^{} \cdot V_{13}^{} \cdot V^{}_{12}}:$
In this scenario, we have
\begin{eqnarray}
&& \tan \theta^{}_{12} = \left|{\cos x - \cos z \sin x \over \cos x
\cos z
-\sin x }\right| \; , \\
&& \sin \theta^{}_{13} =\left|{\sin z \over \sqrt{2}}\right| \; , \\
&& \tan \theta^{}_{23} = \left|{2 \cos z -  \sin z \over \sqrt{2 }
(\cos z + \sin z)} \right|\; ,
\end{eqnarray}
where $\theta^{}_{13}$ and $\theta^{}_{23}$ have the following
correlation:
\begin{eqnarray}
\sin \theta^{}_{13} = {\sqrt{2} - \tan \theta^{}_{23} \over \sqrt{4
\tan^2 \theta^{}_{23}-2\sqrt{2}\tan \theta^{}_{23} +5}} \; .
\end{eqnarray}
\begin{figure}[h!]
\includegraphics[width=2.81in]{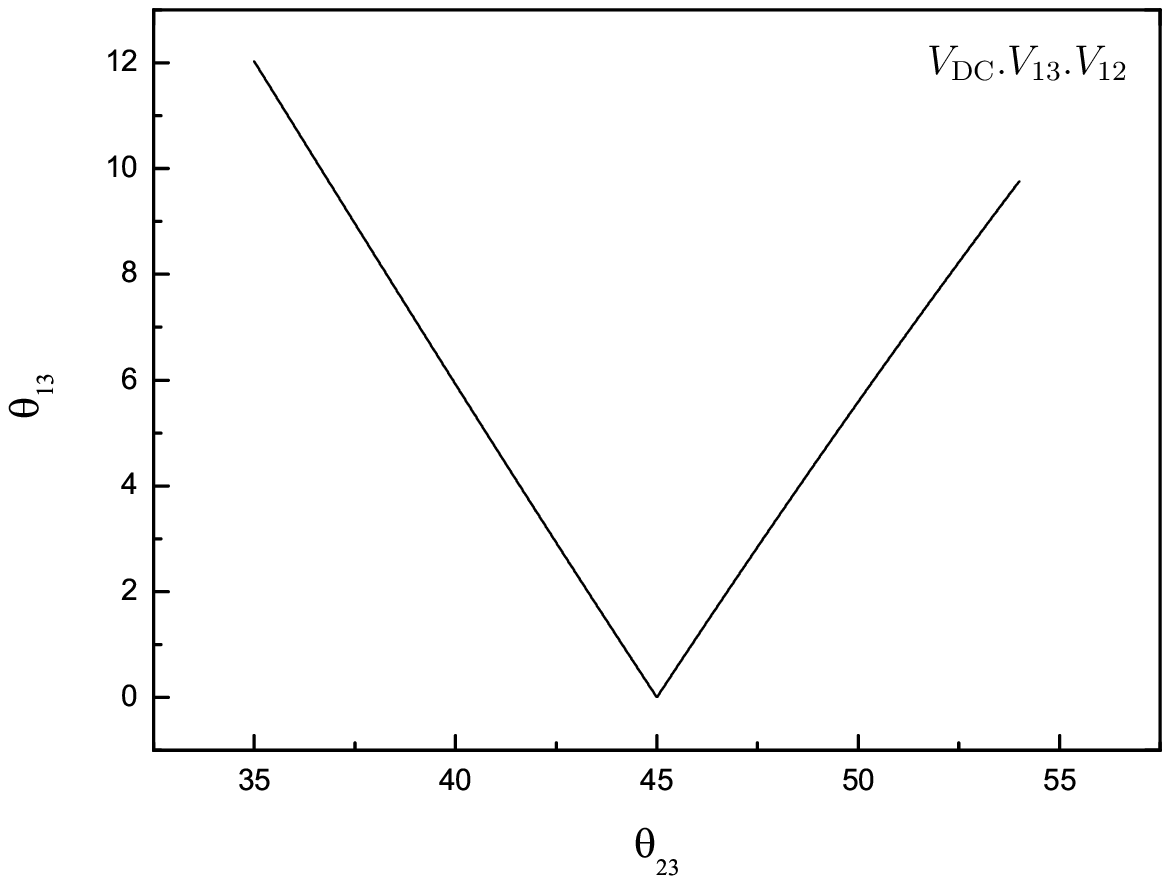}
\includegraphics[width=2.87in]{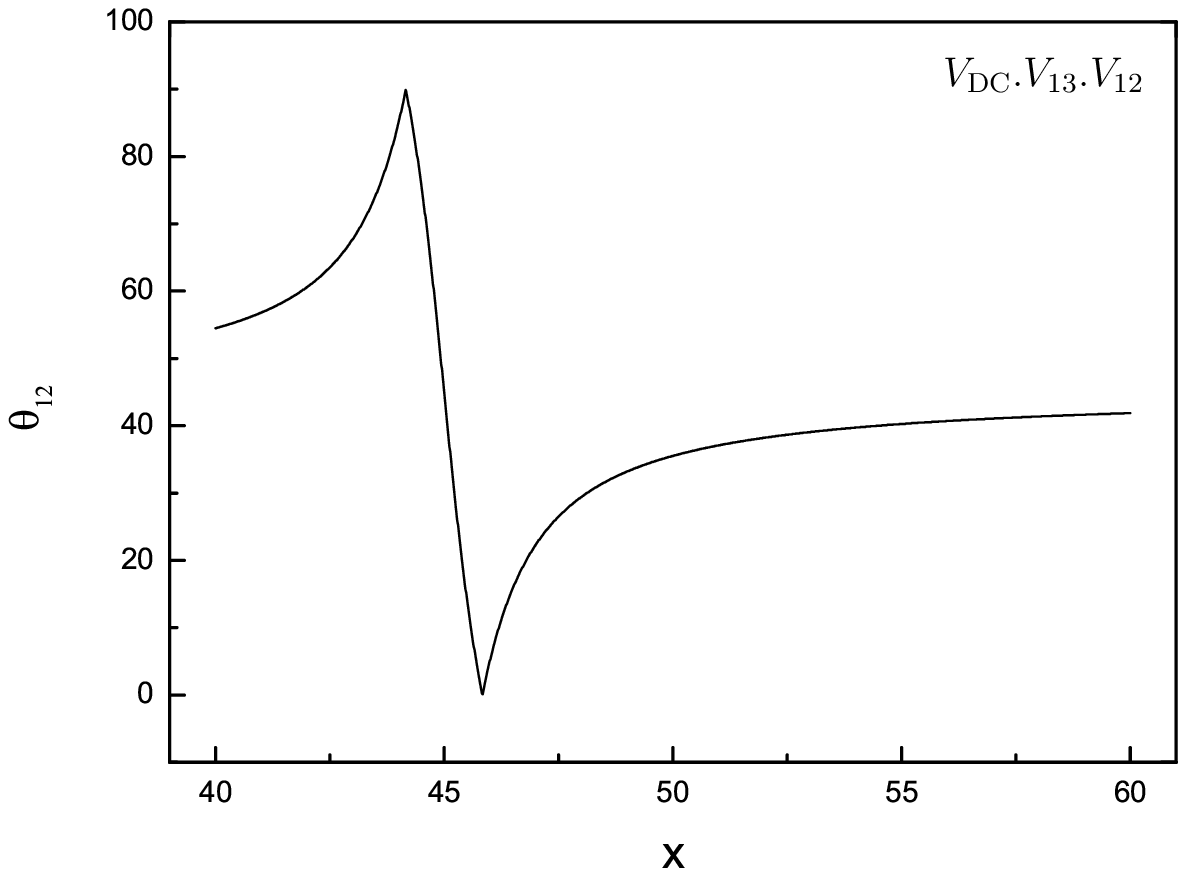}
\caption{Case VDCrr1. $\theta_{13}$ as a function of $\theta_{22}$
(left panel) and $\theta_{12}$ as a function of $y$ (right panel).}
\label{fig:DCrr1}
\end{figure}
We plot in the left panel of Fig. \ref{fig:DCrr1} $\theta^{}_{13}$
as function of $\theta^{}_{23}$. In this scenario the possible
parameter range of $\theta^{}_{13}$ is $[3.5^\circ, ~ 16.5^\circ ]$.
Big $\theta^{}_{13}$ is permitted in this scenario. If future
reactor neutrino oscillation experiment constraint $\theta^{}_{13}$
lying below $3^\circ$, this scenario can be excluded.  We also plot
$\theta^{}_{12}$ as function of $x$, given $\theta^{}$ it's best fit
value $9.7^\circ$. We may obtain appropriate $\theta^{}_{}$ with big
rotation angle $x$.

\item $\mathbf{VDCrr2=  V_{\rm DC}^{} \cdot V_{23}^{} \cdot
V^{}_{12}}:$ The phenomenology of this scenario was already studied
in paper \cite{xing}. We won't repeat them here.

\item $\mathbf{VDCrr3=  V_{\rm DC}^{} \cdot V_{23}^{} \cdot
V^{}_{13}:}$ We may derive the following equations in this scenario
\begin{eqnarray}
&& \tan \theta^{}_{12} =\left| { \cos y \over \cos z - \sin y \sin
z} \right|\; ,
\\
&& \sin \theta^{}_{13} =\left| { \sin z + \cos z \sin y \over
\sqrt{2} } \right|
\; , \\
&& \tan \theta^{}_{23} =\left|{ 2 \cos y \cos z +\cos z \sin y -
\sin z \over \sqrt{2} (\cos y \cos z - \cos z \sin y + \sin z )
}\right| \; .
\end{eqnarray}
And we have
\begin{eqnarray}
&&\sin z = { (1 + \sqrt{2} \tan \theta^{}_{23}) \sin \theta^{}_{13}
+ \tan \theta^{}_{12} (\sqrt{2}- \tan \theta^{}_{23}) \over \sqrt{2}
+ 2 \tan \theta^{}_{23} + \sin \theta^{}_{13} \tan \theta^{}_{12} (2
-\sqrt{2}\tan \theta^{}_{23})} \; ,\\
&& \sin y = \sqrt{\cot^2 \theta^{}_{12} + 2 \sin^2 \theta_{13} -1
\over 1 + \cot^2 \theta^{}_{12} } \; .
\end{eqnarray}
\begin{figure}[h!]
\includegraphics[width=3.5in]{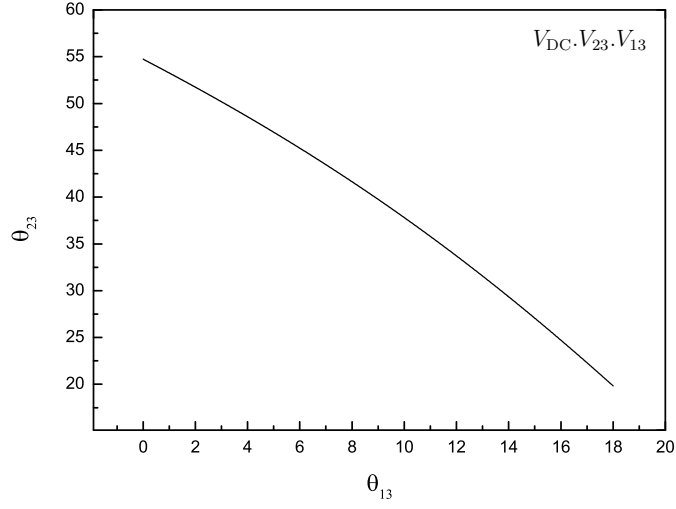}
\caption{Case VDCrr3. $\theta_{23}$ as a function of $\theta_{13}$
.} \label{fig:DCrr3}
\end{figure}
Given these correlations, we may plot $\theta^{}_{23}$ as function
of $\theta^{}_{13}$ by choosing $\theta^{}_{12}$ its best fit value.
From Fig. \ref{fig:DCrr3}, we may read parameter space of
$\theta^{}_{13}$ by setting $\theta^{}_{23}$ changing in it's
$3\sigma$ range, which is $[0.9^\circ,~11.1^\circ]$. It's an
excellent parameter space, covering T2K's best fit value for both
normal hierarchy case and inverted hierarchy case.

\item $\mathbf{VDC\ell \ell 1 = V_{13}^{} \cdot V^{}_{12} \cdot V_{\rm
DC}^{} :}$ In this scenario, we have
\begin{eqnarray}
&&\tan \theta^{}_{12} =\left| { \sqrt{6} \cos x \cos z + \sqrt{2}
\sin x \cos z + 2 \sin z \over \sqrt{6} \cos x \cos z - \sqrt{2}
\sin x
\cos z - 2 \sin z}\right| \; , \\
&& \sin \theta^{}_{13} =\left|{ \sqrt{2 } \sin x \cos z - \sin z
\over
\sqrt{3}}\right| \; , \\
&& \tan \theta^{}_{23}  =\left| { \sqrt{2} \cos x \over \cos z +
\sqrt{2} \sin x \sin z }\right| \; .
\end{eqnarray}
Here, $x$ and $y$ may be expressed as functions of $\theta^{}_{ij}$
\begin{eqnarray}
&&\sin x = \sqrt{{2 \cot^2 \theta^{}_{23} + 3 \sin^2
\theta^{}_{13}-1 \over 2(1 + \cot^2 \theta^{}_{23})}} \; , \\
&&\sin z = {\tan \theta^{}_{23} ( \tan \theta^{}_{12} -1 ) - \sin
\theta^{}_{13} (1 + \tan \theta^{}_{12} ) \over 3(1 + \tan
\theta^{}_{12}) + \sqrt{6} \sin \theta^{}_{13} \tan \theta^{}_{23}
(1- \tan \theta^{}_{12})} \; .
\end{eqnarray}
\begin{figure}[h!]
\includegraphics[width=3.5in]{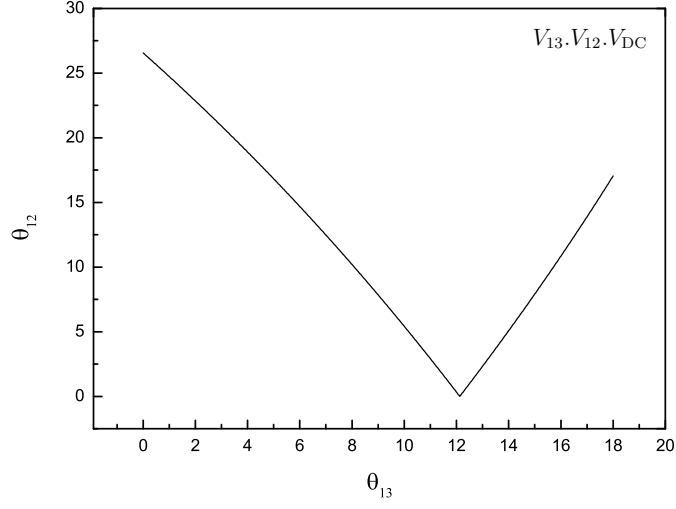}
\caption{Case VDCll1. $\theta_{13}$ as a function of $\theta_{12}$
.} \label{fig:DCll1}
\end{figure}
We plot in Fig. \ref{fig:DCll1} $\theta^{}_{23}$ as function of
$\theta^{}_{13}$, by setting $\theta^{}_{23}$ its best fit value.
It's clear that we can not get the appropriate $\theta^{}_{12}$ when
$\theta^{}_{13}$ changes in the range $[0^\circ,~18^\circ]$ in this
case. Such that it's excluded. However, if we let $\theta^{}_{23}$
changing in its $3\sigma$ range, a narrow  appropriate parameter
space of $\theta^{}_{12}$ can be obtained. In short, this scenario
doesn't work for the best value.

\item
$\mathbf{VDC\ell \ell 2 = V_{23}^{} \cdot V^{}_{12} \cdot V_{\rm
DC}^{}: }$ In this case we have
\begin{eqnarray}
&&\tan \theta^{}_{12} = \left|{\sqrt{3} \cos x + \sin x \over
\sqrt{3} \cos
x -\sin x } \right|\; , \\
&&\sin \theta^{}_{13} = \left|{\sqrt{2 \over 3} \sin x }\right| \; , \\
&& \tan \theta^{}_{23} = \left|{ \sqrt{2} \cos x \cos y + \sin y
\over \cos y - \sqrt{2} \cos x \sin y} \right|\; ,
\end{eqnarray}
The correlation between $\theta^{}_{13}$ and $\theta^{}_{12}$ is
\begin{eqnarray}
\sin \theta^{}_{13} = {1 - \tan \theta^{}_{12} \over \sqrt{2 + 2
\tan^2 \theta^{}_{12}-2\tan \theta^{}_{12}}} \; .
\end{eqnarray}
We plot in the left panel of Fig. \ref{fig:DCll2}, $\theta^{}_{13}$
as function of $\theta^{}_{12}$ and in the right panel of Fig.
\ref{fig:DCll2} $\theta^{}_{23}$ as function of y. We find that
$\theta^{}_{13}$ can only changes only in the range $[12.5^\circ, ~
21.3^\circ]$. This range is excluded by the best fit value, but
allowed by the T2K's result. If future reactor neutrino oscillation
experiments negate the T2K's result, this scenario will be excluded
definitely.

\item
$\mathbf{VDC\ell \ell 3 = V_{23}^{} \cdot V^{}_{13} \cdot V_{\rm
DC}^{}: }$  The following equations can be obtained from this
scenario
\begin{eqnarray}
&&\tan \theta^{}_{12} = \left|{ \sqrt{3} \cos z + \sqrt{2} \sin z
\over
\sqrt{3} \cos z - \sqrt{2 } \sin z}\right| \; , \\
&& \sin \theta^{}_{13} = \left|{\sin z \over \sqrt{3}}\right| \; , \\
&& \tan \theta^{}_{23} = \left| { \sqrt{2} \cos y + \cos z \sin y
\over \cos y \cos z - \sqrt{2} \sin y} \right|\; .
\end{eqnarray}
We have
\begin{eqnarray}
\sin \theta^{}_{13} = { 1 - \tan \theta^{}_{12} \over \sqrt{5 \tan^2
\theta^{}_{12}-2 \tan \theta^{}_{12} +5}} \; .
\end{eqnarray}
We plot in Fig.\ref{fig:DCll3} $\theta^{}_{13}$ as function of
$\theta^{}_{12}$ and  $\theta^{}_{23}$ as function of y. It's clear
from the figure that $\theta^{}_{13}$ can only change in the space
$[6.4^\circ, ~10.5^\circ]$ in this scenario. We can get appropriate
$\theta^{}_{23}$ with big rotation angle $y$, which can be the
result of certain flavor symmetry.

\begin{figure}[h!]
\includegraphics[width=2.81in]{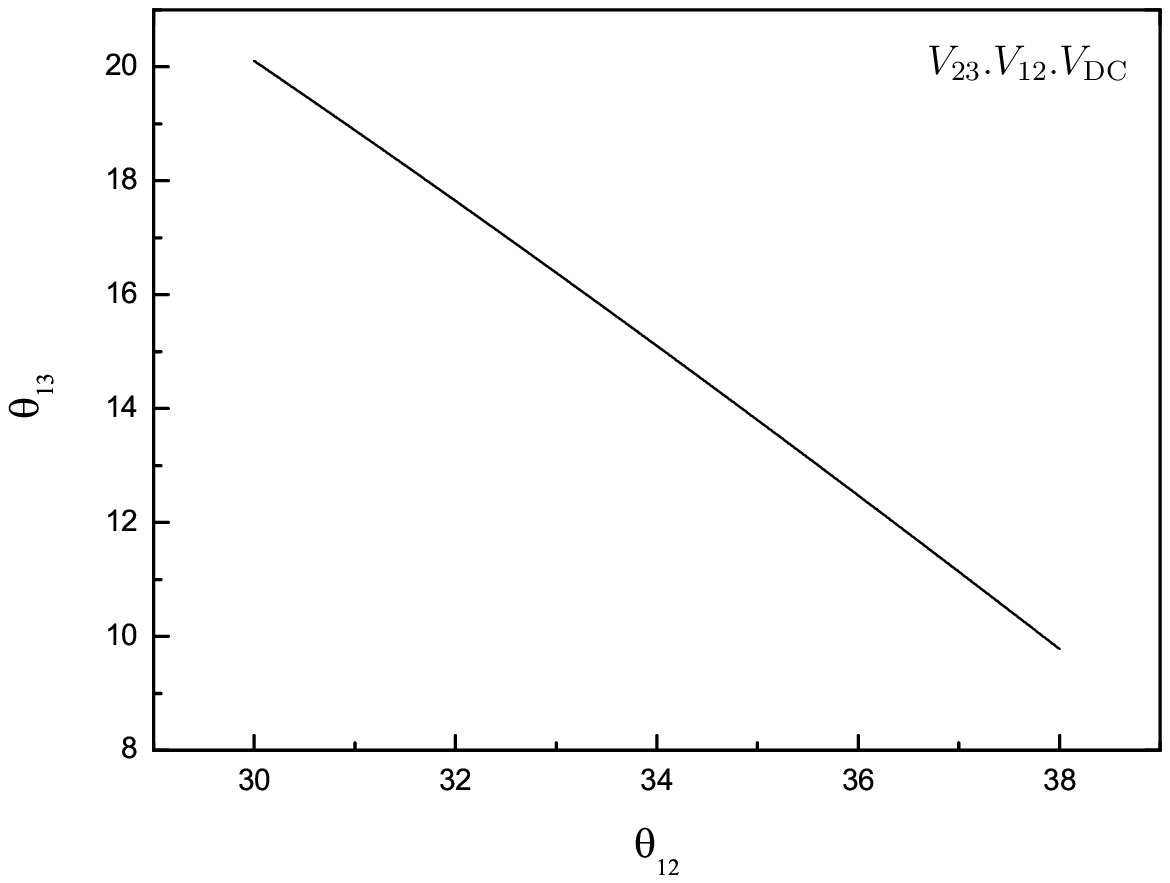}
\includegraphics[width=2.81in]{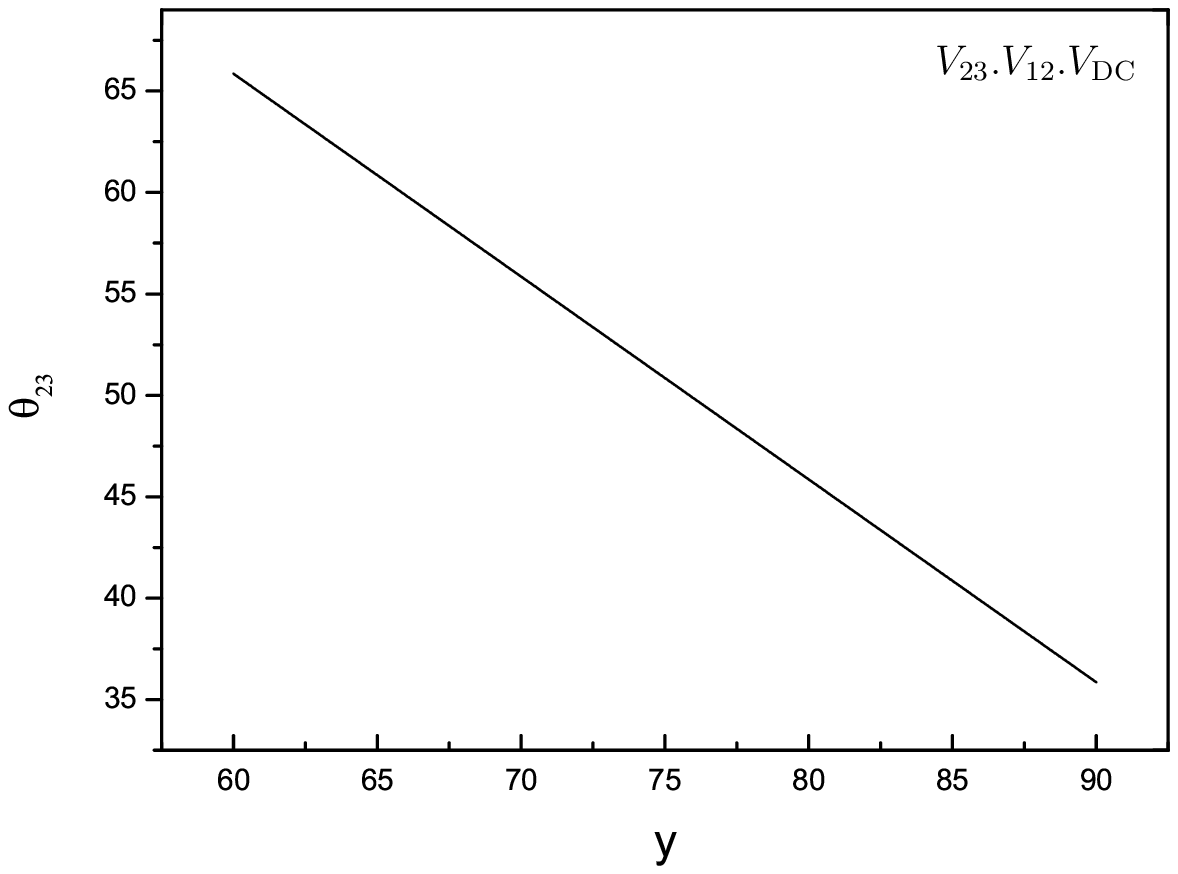}
\caption{Case VDCll2. $\theta_{13}$ as a function of $\theta_{12}$
(left panel) and $\theta_{23}$ as a function of $y$ (right panel).}
\label{fig:DCll2}
\end{figure}

\begin{figure}[h!]
\includegraphics[width=2.81in]{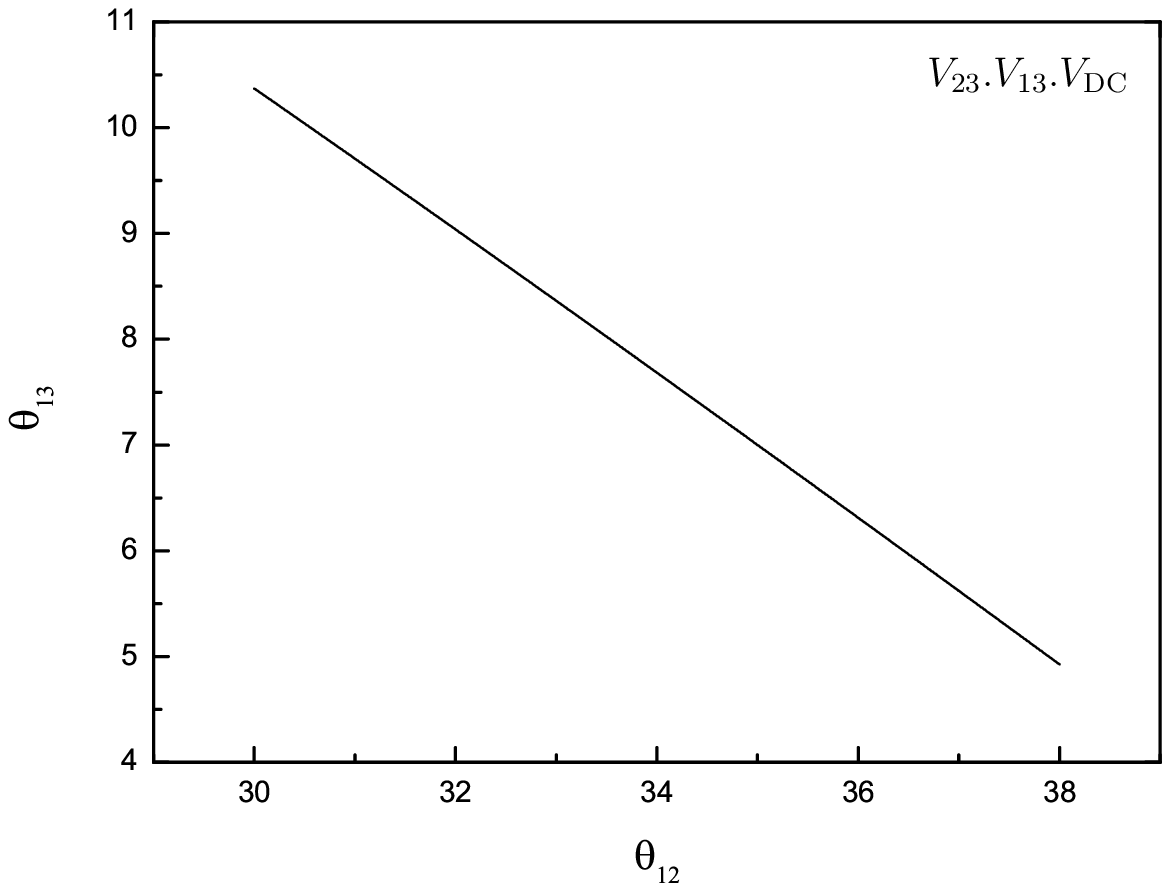}
\includegraphics[width=2.81in]{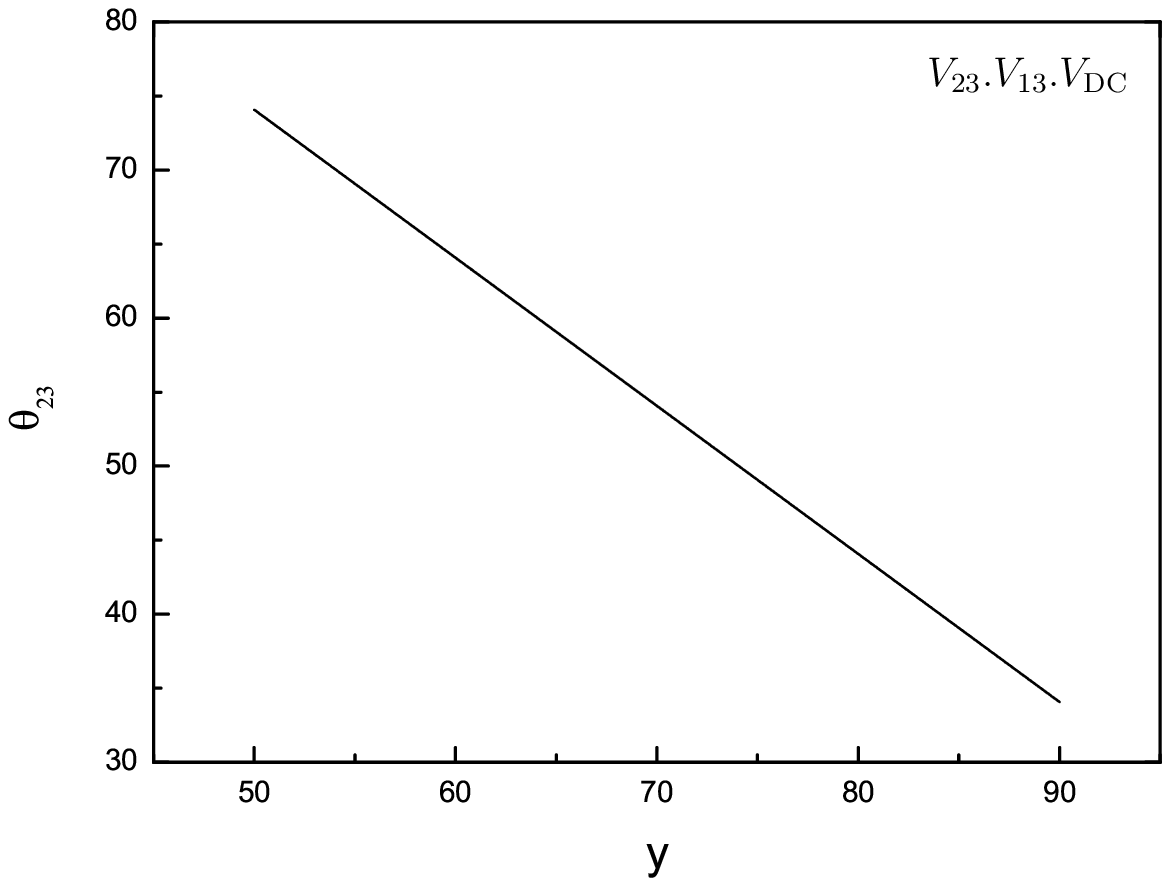}
\caption{Case VDCll3. $\theta_{13}$ as a function of $\theta_{12}$
(left panel) and $\theta_{23}$ as a function of $y$ (right panel).}
\label{fig:DCll3}
\end{figure}

\end{itemize}


\section{Summary}\label{section:summary}
Among the knowns and unknowns of neutrino physics, the nonzero
smallest lepton  mixing angle $\theta_{13}$ has received a lot of
attentions. The studying and  measurement of $\theta_{13}$ is
important to enrich our understanding of neutrino properties.
Inspired by the recent T2K result of a relatively large
$\theta^{}_{13}$, We investigated  some general modifications, which
may minimal modifications or non-minimal modifications to the three
well-known neutrino mixing pattern: Bimaximal, Tri-bimaximal and
Democratic mixing patterns. Some non-trivial correlations between
$\theta^{}_{12}$, $\theta^{}_{23}$ and $\theta^{}_{13}$ were
obtained. We investigated constraints on the  $\theta^{}_{13}$ by
these correlations. Some modifications are already excluded by the
current neutrino oscillation data, while the others give their
predications on $\theta^{}_{13}$. Future neutrino oscillation
experiments may confirm or negate these models. Since
$\theta^{}_{13}$ is not so small, investigating leptonic
CP-violating effects become relevant, but beyond the scope of this
preliminary work. We will study this interesting and important topic
somewhere else.

\begin{acknowledgments}
One of the authors (Y.Zheng) would like to thank Prof. B. Q. Ma for
his hospitality during her stay in Peking University. This work is
partially supported by Chinese PostDoc Foundation (Grants
No.~45210148-0172)(W.Chao) and by Peking University Visiting Scholar
Program for Graduate Students(Y.Zheng).

\end{acknowledgments}

\end{document}